\documentclass{article}

\usepackage{PRIMEarxiv}

\usepackage[utf8]{inputenc} 
\usepackage[T1]{fontenc}    
\usepackage{hyperref}       
\usepackage{url}            
\usepackage{booktabs}       
\usepackage{amsfonts}       
\usepackage{nicefrac}       
\usepackage{microtype}      
\usepackage{lipsum}
\usepackage{fancyhdr}       
\usepackage{graphicx}       
\graphicspath{{media/}}     

\title{The direct democracy paradox: Microtargeting and issue ownership in Swiss online political ads
}

\author{
  Arthur Capozzi \\
  \texttt{Computational Social Science} \\
  \texttt{ETH Z\"{u}rich} \\
  \texttt{Stampfenbachstrasse 48}, \texttt{8006}, \texttt{Z\"{u}rich}, \\
  \texttt{Switzerland}
}

\begin{document}
\maketitle

\begin{abstract}
Political advertising on social media has fundamentally reshaped democratic deliberation, playing a central role in electoral campaigns and propaganda. However, its systemic impact remains largely theoretical or unexplored, raising critical concerns about institutional fairness and algorithmic transparency. 
This paper provides the first data-driven analysis of the relationship between direct democracy and political advertising on social media, leveraging a novel dataset of 40,000 political ads published on Meta in Switzerland between 2021 and 2025.
Switzerland's system of direct democracy, characterized by frequent referenda, provides an ideal context for examining this relationship beyond standard electoral cycles.
The results reveal the sheer scale of digital campaigning, with 560 million impressions targeting 5.6 million voters, and suggest that greater exposure to ``pro-Yes'' advertising significantly correlates with referendum approval outcomes.
Demographic microtargeting analysis suggests partisan strategies: Centrist and right-wing parties predominantly target older men, whereas left-wing parties focus on young women.
Regarding textual content, a clear pattern of ``talking past each other'' is identified; in line with the issue ownership theory, parties avoid debating shared issues, preferring to promote exclusively owned topics.
Furthermore, the parties' strategies are so distinctive that a machine learning model trained only on audience and topic features can accurately predict the author of an advertisement.
This article highlights how demographic microtargeting, issue divergence, and tailored messages could undermine democratic deliberation, exposing a paradox: Referenda are designed to be the ultimate expression of the popular will, yet they are highly susceptible to invisible algorithmic persuasion.
\end{abstract}


\section{Introduction}

Online political advertising has long been considered controversial and potentially capable of undermining democratic processes, primarily due to its effectiveness in manipulating voters, fueling cynicism, and polarizing the electorate~\cite{JamiesonKathleenHall}. This reality stands in stark contrast to the principles of active democracy, which is founded on the idea that citizens are well-informed and capable of debating and deciding on complex policy issues as part of shared public deliberation~\cite{BLONDELJEAN}. The threat to this public sphere is significantly amplified by the advent of microtargeting, which uses vast datasets to tailor persuasive messages to narrow, psychologically profiled segments of the electorate~\cite{ZuiderveenBorgesiusFrederik}. In fact, multiple studies have found that such personalized political ads are more effective at persuading than generic, non-personalized messages~\cite{SimchonEdwards, TappinWittenberg}, while new generative AI technologies promise to render these microtargeting campaigns even more effective~\cite{SimchonEdwards}.
Although the impact of these technologies on candidate elections is well documented~\cite{CapozziThin, BarFeuerriegel}, their systemic impact on policy-making and direct legislation remains largely unexplored.
Does algorithmic microtargeting simply encourage voter participation, or does it fundamentally fragment the debate on key issues? This gap is critical because the risks are amplified in direct democratic systems.
Here, a dangerous paradox emerges: citizens operate under the impression that they are making sovereign decisions, yet their preferences may actually be steered by opaque, tailored messaging. If the electorate feels empowered to decide but is unknowingly steered by algorithms, is the ``will of the people'' still authentic?

To answer this question, Switzerland offers a unique laboratory. Firstly, direct democracy is not just a feature, but a central pillar of the nation's identity. Swiss citizens vote on national issues approximately four times per year, participating in far more referendums and initiatives than citizens of any other country~\cite{Serdült2014}. Secondly, Switzerland is a robust setting for testing the generalizability of cross-cultural findings. Switzerland is a multicultural nation of 9 million people with four official languages, where 41\% of permanent residents have a migration background as of 2024.

This article presents a longitudinal, data-driven analysis of a novel dataset of 40,000 political ads published on Facebook and Instagram in Switzerland from 2021 to 2025. Through a combined study of demographic microtargeting and textual content, this research assesses the scale, scope, and normative impact of opaque online political advertising, while isolating differences and similarities among major political parties and linguistic regions. Beyond the specific case study, the article describes a reproducible research pipeline that can be applied to other contexts and democracies, and provides researchers and policymakers with a toolbox to make the digital public sphere more transparent.

\section{Background}

\subsection{\textbf{Online political advertising}}



The real impact of social media platforms on political opinion formation and voting behavior remains a topic of debate in the scientific literature. For instance, large-scale experiments have demonstrated that algorithmically delivered social messages can measurably increase voter turnout~\cite{bond201261}.
Other experiments have shown that curated search rankings can also potentially shift the preferences of undecided voters~\cite{Epstein}.
The impact of online communication on politically sensitive topics has also been examined. For example, Papakyriakopoulos et al.~\cite{Papakyriakopoulos4} investigated the role of demagoguery in civic engagement on Reddit, while Lee and Hsieh~\cite{LeeYu-Hao} examined participation in online activism, or ``slacktivism'', in donation campaigns.
Memes and other new forms of communication have also been studied as a mean of political propaganda on social media~\cite{WilliamsNagel, Mejova, Leon}.

This online political context intersects with the underlying business model of social media platforms, which is predominantly based on advertising~\cite{ANDERSON201541}. The economic logic incentivizes the continuous algorithmic optimization of content delivery systems to maximize user engagement and, consequently, advertising revenue~\cite{O’Reilly_Strauss_Mazzucato_2024, RanDezhi}.
The exploitation of these highly optimized, profit-driven systems for the dissemination of political messages is, on the one hand, a source of contemporary public debate and academic research~\cite{DommettKatharine, AliSapiezynski}.
On the other hand, the effectiveness of recommendation systems has also made online ads much more effective~\cite{ChuVliegenthart} and, as a result, led to a significant increase in their use~\cite{Sosnovik}.

In this context, Meta's platforms, Facebook and Instagram, are among the most effective at influencing public opinion~\cite{KarpfDavid} and accurately identifying users' demographic information~\cite{DanielaPerrotta}.
Madrigal~\cite{madrigal2017facebook} argues that although Facebook's potential impact on elections had been evident for years, it reached its peak during the 2016 presidential election, with \$1.4 billion spent on online ads during the campaign.
However, hundreds of millions of dollars of dark ads paid for by foreign governments to influence elections should be added to this number~\cite{madrigal2017facebook}.
In this regard, it has been proven that Russia attempted to manipulate the 2016 U.S. presidential election by running political ads on Facebook~\cite{entous2017russian, EmilioFerrara}.
Matias et al.~\cite{MatiasNathan} designed a software-supported approach for auditing, which uses coordinated volunteers to analyze political advertising policies enacted by Facebook and Google during the 2018 midterm U.S. election.

During the 2021 German federal elections, a positive correlation was identified between social media advertising and successful election outcomes for candidates~\cite{BarFeuerriegel}. In addition, Bär et al.~\cite{PierriGianmarco} found that Meta’s algorithm prioritized advertisements from populist parties during the same election campaign.
Other studies on Facebook political ads include anti-vaccination ads during COVID-19~\cite{MejovaKalimeri}, populist advertising for the European Parliamentary election in 2019~\cite{CapozziThin}, Italian political propaganda on the migration issue~\cite{capozzi2021clandestino, capozzi2020facebook}, far-right misinformation spread by the populist Spanish party VOX~\cite{CanoLorena}.
Many studies examined the spread and impact of political ads in Switzerland, with the earliest focusing on traditional newspapers~\cite{GerberBühlmann, ZumofenGuillaume}.
Other studies have focused on referendum campaigns shifting from newspapers to social media~\cite{FischerMichaela}, or politicians' use of social media~\cite{ReveilhacMorselli}.

\subsection{\textbf{Switzerland and direct democracy}}

Even before the advent of social media, the efficacy of direct democracy was questioned in academic literature, with some analyzes finding it to be ineffective or even counterproductive in practice. Critics often highlight the potential for a ``tyranny of the majority'', the influence of wealthy special interest groups, and the risk posed by citizen incompetence regarding complex policy issues~\cite{warner1995direct, gerber2011populist}.
However, Switzerland is often cited as an exception to this criticism, with its political system considered a model of success~\cite{BrunoFrey}. The Swiss model is not a simple and universally applicable template; rather, it has developed within a specific and unique historical and institutional setting. It evolved ``bottom-up'' from cantonal traditions, not being imposed by a central authority~\cite{Linder2010}.
This high degree of direct democracy and participation is argued to have numerous positive social and economic effects, including empowering citizens, making procedures more appealing, and imposing fiscal discipline on public spending~\cite{BrunoFrey, STUTZERALOISBRUNO}. Further research connects direct democratic rights directly to improved economic performance~\cite{FeldLarsSaviozMarcel} and higher levels of subjective well-being, or happiness, among citizens~\cite{FreyBrunoStutzerAlois2000}.

Even within the Swiss context, the positive portrayal of direct democracy has its detractors. Some scholars have highlighted problematic aspects, including ``voting fatigue'' caused by the large number of decisions, the complexity of certain ballot measures, and the system's ability to impede social and political reforms~\cite{BLONDELJEAN}. Furthermore, the ``double majority'' rule, which requires a majority of both citizens and cantons, has been criticized for giving disproportionate power to small, rural cantons. This rule enables controversial initiatives to pass despite lacking broad popular support in major population centers~\cite{Papadopoulos2001}.

In order to understand the deep cultural roots of direct democracy in Swiss institutions and society, it is necessary to start with the very foundations of the Swiss Confederation.
The modern federal state was established in 1848 as a direct result of a brief civil war, known as the Sonderbund War. The new state faced the challenges of integrating minorities and managing the country's deep ``cross-cutting cleavages''~\cite{Linder2010}. Switzerland was and remains a mosaic:

\textit{Linguistic.} Since unification, there have been four national languages: German, French, Italian, and Romansh. In 2023, 61.4\% of the population is native German speaker; 22.6\% are native French speakers; 7.7\% are native Italian speakers; and 0.5\% are native Romansh speakers~\cite{FSOLang}.

\textit{Religious.} According to the Swiss National Survey~\cite{FSOrelsurvey}, 35.6\% of the population was religiously unaffiliated in 2023. The remaining population was divided as follows: 30.7\% Catholic, 19.5\% Swiss Protestant, 5.8\% other Christian denominations, and 6\% Muslim.

\textit{Geographic.} Although rural municipalities cover 57\% of the country’s total area, in 2025 they are home to only 35\% of the population. In contrast, urban areas account for just 17\% of the total area but are home to 65\% of the population and 76\% of the workforce~\cite{FSOGeo}.

\textit{Economic.} Cantons with major urban centers, such as Zurich and Basel-Stadt, have an exceptionally high economic output compared to other cantons. For instance, the nominal GDP per capita in Basel-Stadt was about CHF 210,000 in 2022, compared to the Swiss national average of around CHF 90,000 and the canton of Uri's average of CHF 58,000~\cite{FSO_GDP}.

The federal state's guarantee of municipal and cantonal autonomy with regard to taxation, education, and infrastructure has played an important role in preserving cohesion~\cite{Sager_Zollinger_2011}.
Alongside this federal structure, the system of direct democracy has been the most important political compromise made to maintain cohesion in diversity since the unification in 1848.
For this reason, Swiss direct democracy has often been interpreted by scholars as a mechanism for minority protection and power-sharing, specifically designed to integrate the country's diverse segments~\cite{Linder2010, Disset2010}.
The legal instruments available to citizens and enshrined in the Swiss Federal Constitution include the compulsory referendum, non-compulsory (facultative) referendum, and the popular initiative~\cite{Kriesi2005, Papadopoulos2001}.

Swiss direct democracy operates within a historically highly fragmented, multipolar party system that reflects the nation's underlying divisions.
In 2025, the Federal Assembly remains highly multipolar, with no single party holding a majority.
Following the 2023 federal elections, the national-conservative SVP (Swiss People's Party) affirmed its position as the largest party with 27.9\% of the vote. The center-left SP (Social Democratic Party) followed at 18.3\%. The political center is split between the liberal, center-right FDP (FDP.The Liberals) at 14.3\% and the Christian democratic Die Mitte (The Center) at 14.1\%. The environmentalist bloc, which saw historic gains in 2019, is divided between the left-leaning GRÜNE (Green Party) at 9.8\% and the centrist GLP (Green Liberal Party) at 7.6\%~\cite{FSO_Elections2023}.
Finally, it should be mentioned that, out of a population of approximately 9.1 million in 2025, only about 5.6 million people are eligible to vote.







\section{Results}

The findings presented in this section are divided into three subsections: demographic microtargeting (RQ1), impact on direct democratic votes (RQ2), and strategic ownership of issues during federal elections (RQ3). The research questions are the following:
\vspace{14pt}
{
\setlength{\leftmargini}{0.1em}  
\setlength{\leftmarginii}{0.5em} 
\begin{itemize}
    \item[]\textbf{RQ1 Demographic microtargeting}
    \begin{itemize}
        \item[]\textbf{a}: Do parties target specific users based on demographics?
        \item[]\textbf{b}: Are there regional-specific demographics?
    \end{itemize}
    \item[]\textbf{RQ2 Referenda}
    \begin{itemize}
        \item[]\textbf{a}: Are ads used intensively during popular referenda?
        \item[]\textbf{b}: To what extent can ads influence the outcome of a referendum?
    \end{itemize}
    \item[]\textbf{RQ3 Federal elections}
    \begin{itemize}
        \item[]\textbf{a}: Which issues do the parties promote the most, and how?
        \item[]\textbf{b}: Are the content and the microtargeting of an ad reliable predictors of its author?
    \end{itemize}
\end{itemize}
}

\subsection{\textbf{Microtargeting}}

\begin{figure*}[tb]
    \centering
    \includegraphics[width=.99\textwidth]{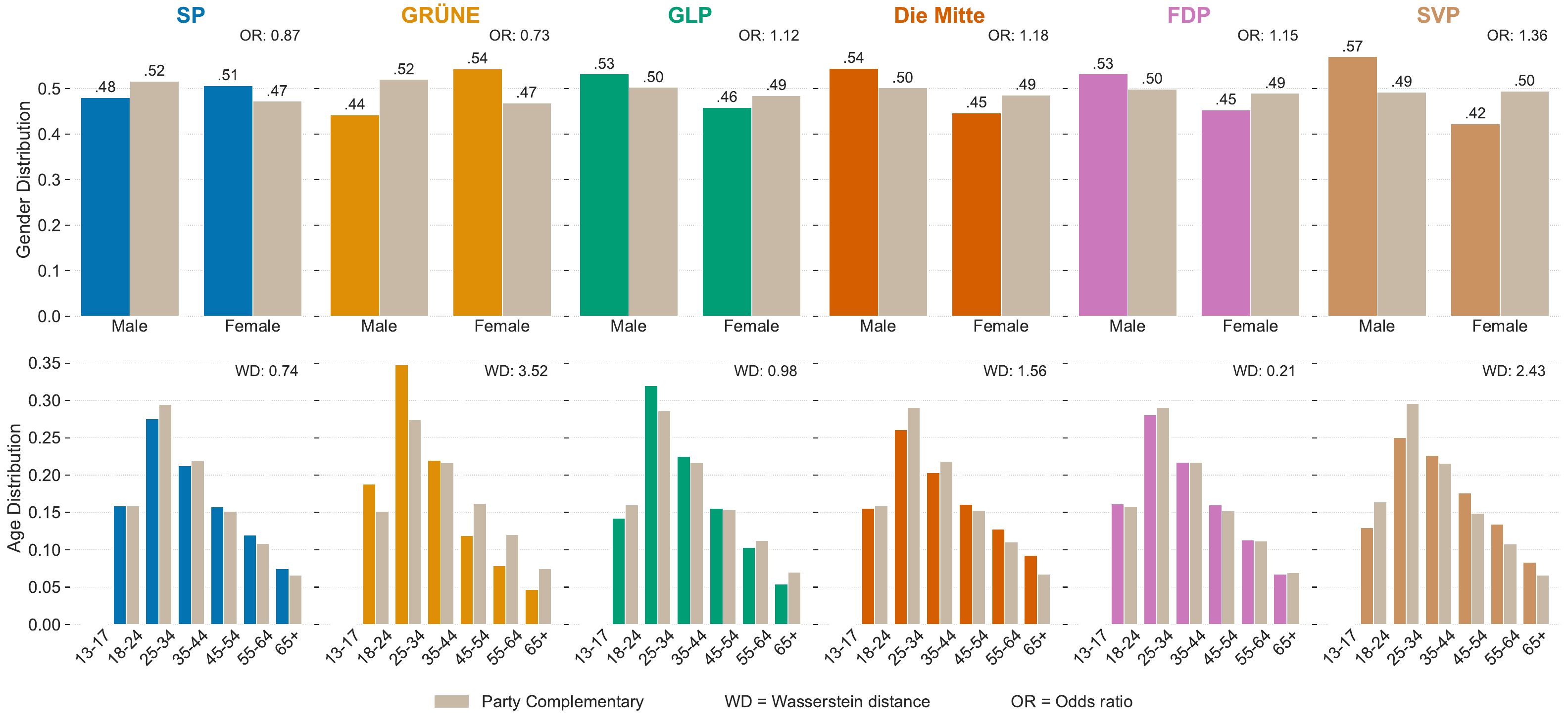}
    \caption{\textbf{Comparison of demographic reach by age and gender.} Each party's audience is compared against the distribution of all other ads (Party Complementary). Gender bias is measured using the male-to-female Odds Ratio (OR), and age distribution shifts are measured using the Wasserstein Distance (WD).}
  \label{fig:demographic_distribution}
\end{figure*}

Figure~\ref{fig:demographic_distribution} compares each party's audience demographics to the ``party complementary'' profile, which consists of all advertisements not associated with that party. Since \textbf{RQ1.a} does not concern regional differences, ads from all languages are combined for this analysis.

Overall, centrist and right-wing parties are more oriented towards a male audience, particularly the SVP ($+36\%$). On the other hand, the GRÜNE and the SP are more female-oriented, with the GRÜNE at $+27\%$ and the SP at $+13\%$. 
In terms of age reach, the SVP and, to a lesser extent, Die Mitte tend to attract an older audience, whereas the GRÜNE and the GLP primarily attract people in the 25–34 age group. Furthermore, the GRÜNE is the only party that extensively reaches the 18–24 age group, yet it underrepresents the over-45 age group more than any other party.

\begin{figure}[tb]
    \centering
    \includegraphics[width=.5\columnwidth]{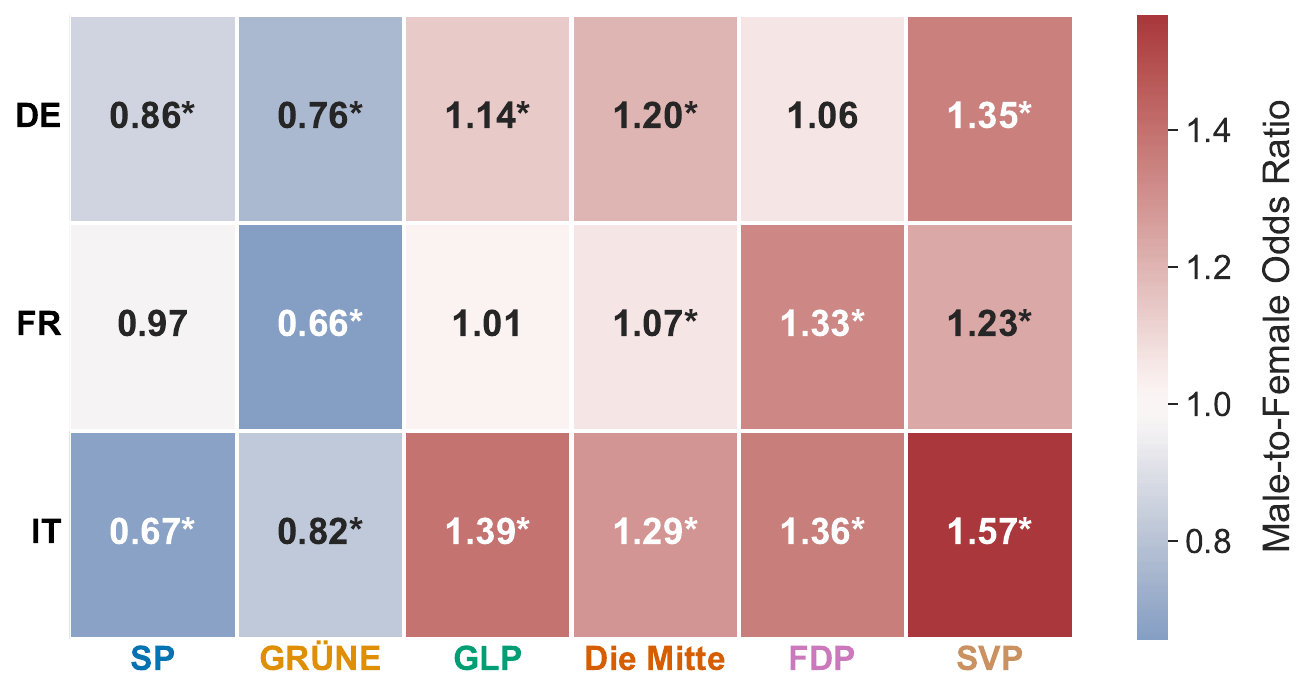}
    \caption{\textbf{Male-to-female targeting bias across linguistic regions.} Odds ratios comparing the gender distribution of each party's ads against the complementary set. Values are bootstrapped means, with 1.0 representing gender balance, while a value greater than 1.0 indicates a shift toward a more male audience. Significant results, where the 95\% confidence interval does not include the baseline of 1.0, are indicated by an asterisk (*).}
  \label{fig:gender_distribution}
\end{figure}

To answer \textbf{RQ1.b} and evaluate the cantonal differences, the demographic distributions can be stratified by language. Figure~\ref{fig:gender_distribution} shows the male-to-female odds ratio within each party's audience by language.
Each cell (party/language) of the heatmap is computed with bootstrapping. For each of the 100 iterations, the OR is computed on a sample of the party's ads and the complement's ads with replacement. The mean and 95\% confidence interval are computed at the end of the process from the bootstrap distribution. 
While center- and right-leaning parties attract male-dominated audiences in all language regions, the SVP's audience in Ticino, the only predominantly Italian-speaking canton, is the most male-oriented.
Other significant regional differences exist. For instance, the GLP and Die Mitte are gender-balanced in the French-speaking cantons and slightly male-oriented in the German-speaking cantons ($+15\%$ and $+20\%$, respectively). However, they are much more male-oriented in Ticino ($+40\%$ and $+28\%$, respectively).
Further to the left on the political spectrum, the situation changes dramatically. For the SP, Ticino is again the least gender-balanced canton, but this time it favors a more female audience ($+32\%$). In German-speaking cantons, the SP reaches a more female audience ($+13\%$), while in French-speaking cantons, the audience is gender balanced. On the other hand the GLP is more gender unbalanced in French-speaking cantons ($+34\%$ women).

\begin{figure*}[tb]
  \centering
  \includegraphics[width=.99\textwidth]{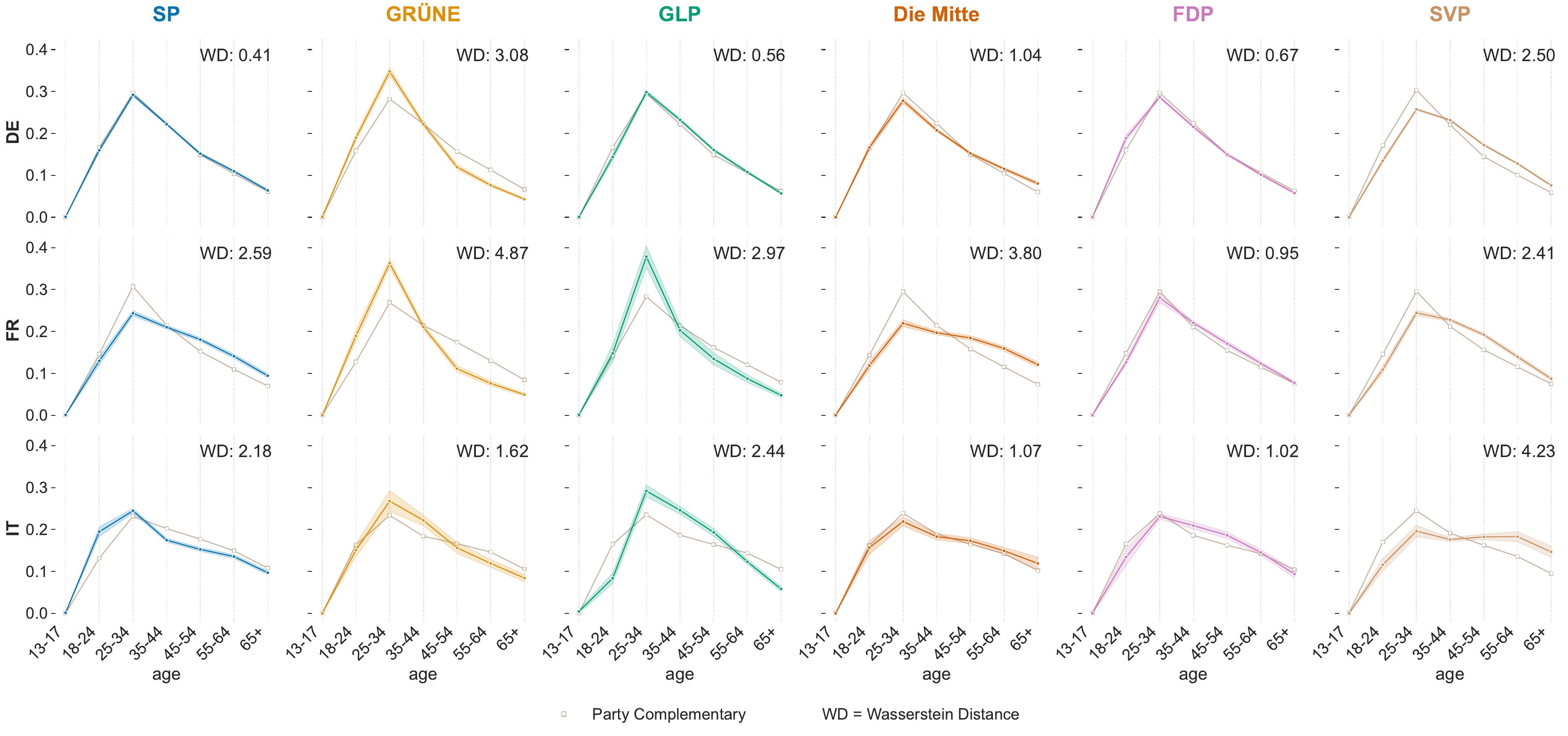}
\caption{\textbf{Language-specific age distributions by party.} Distribution of audience age for each party compared to all other ads (Party Complementary) across German, French, and Italian cantons. Shaded areas represent the standard deviation derived from bootstrapping. Differences in distributions are reported as the Wasserstein Distance (WD).}
  \label{fig:age_distribution}
\end{figure*}

Figure~\ref{fig:age_distribution} shows the age distribution of the audience reached by the parties, compared to the party complementaries, broken down by language group.
The shaded area represents the standard deviation of the audience reached for each age group. The distribution is build using a bootstrapping procedure similar to that described for Figure~\ref{fig:gender_distribution}.
Significant differences between parties within the same canton can be observed, as well as regional variations within parties.
While the SVP focuses on an audience over 45, especially in Ticino, Die Mitte focuses on an older audience in French-speaking cantons only.
The SP, in German-speaking cantons, has a WD of almost 0 (no differences with the complementary party's audience), while in French-speaking cantons, it targets a much more over-45 audience. In Ticino, however, it does the opposite, underrepresenting the over-35 audience and focusing on the 18-24 age range.
The GLP, on the other hand, focuses heavily on the 25-34 age range in French-speaking cantons, while in Ticino, the reach is broader (25-45).
With the exception of the GRÜNE and the SVP, all parties in the German-speaking cantons show fewer differences compared to their counterparts. 
In summary, parties differ in their demographic reach (\textbf{RQ1.a}), with evident internal variations emerging across different linguistic regions (\textbf{RQ1.b}).

\subsection{\textbf{Referenda}}
\label{sec:Referenda}

\begin{figure}[tb]
  \centering
  \includegraphics[width=.8\columnwidth]{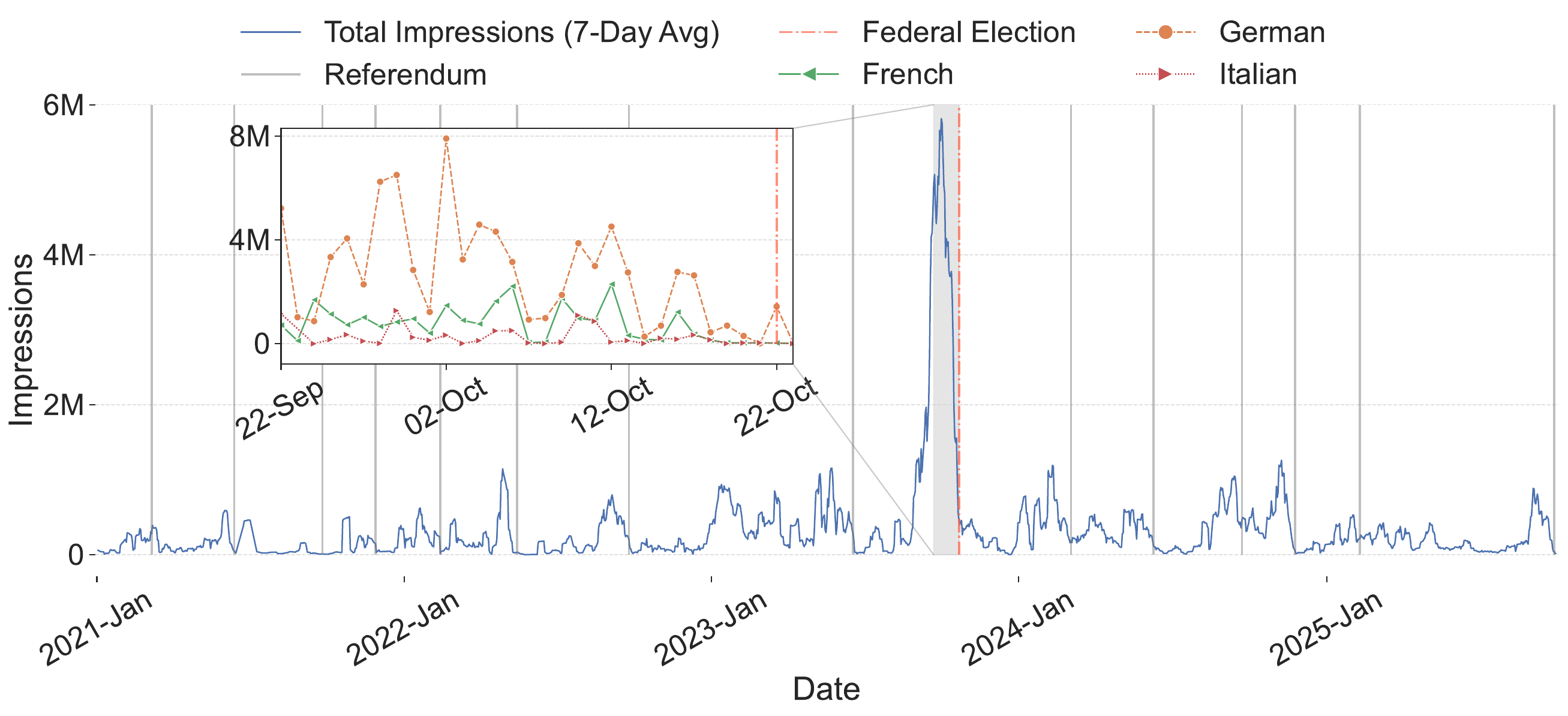}
  \caption{\textbf{Seven-day rolling average of ad impressions.} The inset provides a daily view, disaggregated by language, for the 30-day period preceding the federal election.}
  \label{fig:timeseries_impressions}
\end{figure}

Figure~\ref{fig:timeseries_impressions} shows the seven-day moving average of total political ad views. The gray vertical lines represent referendums, and the inset shows the total number of views per language in the 30 days before the federal elections of October 22, 2023. Clearly, this event caused the most significant increase in views.
At first glance, it is instead difficult to answer \textbf{RQ2.a} and determine whether advertising activity increases before the 14 referendum dates.
As preliminary step, a Wilcoxon signed-rank test is applied to compare ad views and spending in the 30 days before and after each referendum. For completeness, this analysis is conducted separately for German, French, and Italian ads. The results of all the tests (shown in S4 Table) confirm a significant increase in ad views and spending in the 30 days before a referendum ($p < 0.001$ in all cases).

\begin{figure}[ht]
    \centering
    \includegraphics[width=.9\columnwidth]{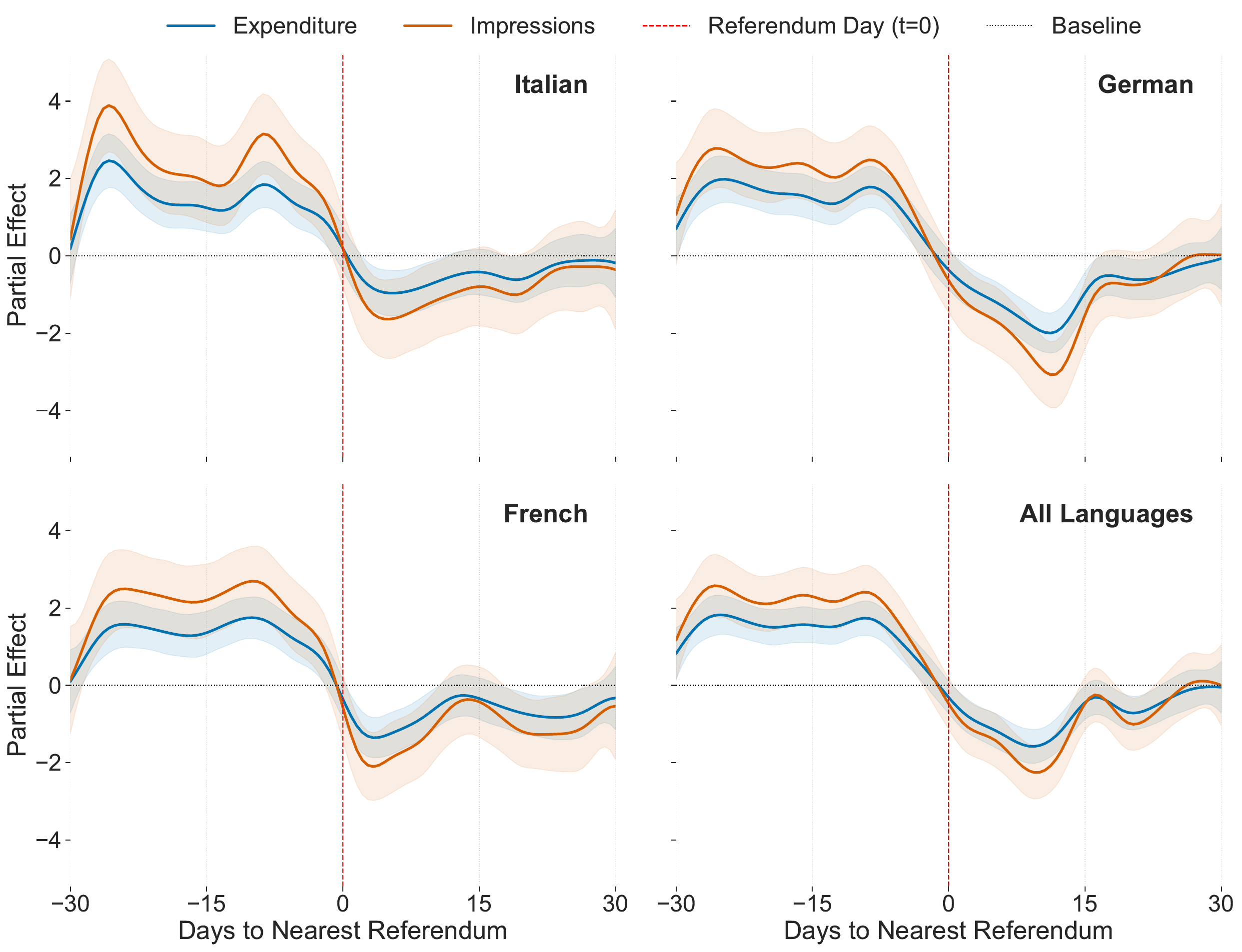}
    \caption{\textbf{Temporal dynamics of advertising intensity.} Partial effects of the ``days to referendum'' variable on log-transformed Expenditure (blue curve) and Impressions (orange curve). The vertical dotted line at $t=0$ marks the day of the vote. The analysis isolates the referendum effect from seasonal trends and is presented for each linguistic region and the national aggregate. Shaded areas represent 95\% confidence intervals.}
  \label{fig:GAM_model}
\end{figure}

Although the Wilcoxon signed-rank test confirms a difference, it does not account for confounding factors, such as seasonality or other temporal patterns. The true impact of the referenda can be isolated by fitting a GAM to predict the daily spending and impressions.
Figure~\ref{fig:GAM_model} shows the partial effects of the proximity of the referendum on the log-transformed expenditure and impressions. The results are consistent across all language groups and suggest a significant increase in ad spending and impressions 30 days before the vote. The 95\% confidence intervals remain entirely above the baseline during this pre-referendum period, and then collapse immediately after the vote at $t=0$. These results confirm that ads are used intensively during the referendum campaign period.

\begin{figure}[ht]
  \centering
  \includegraphics[width=.5\columnwidth]{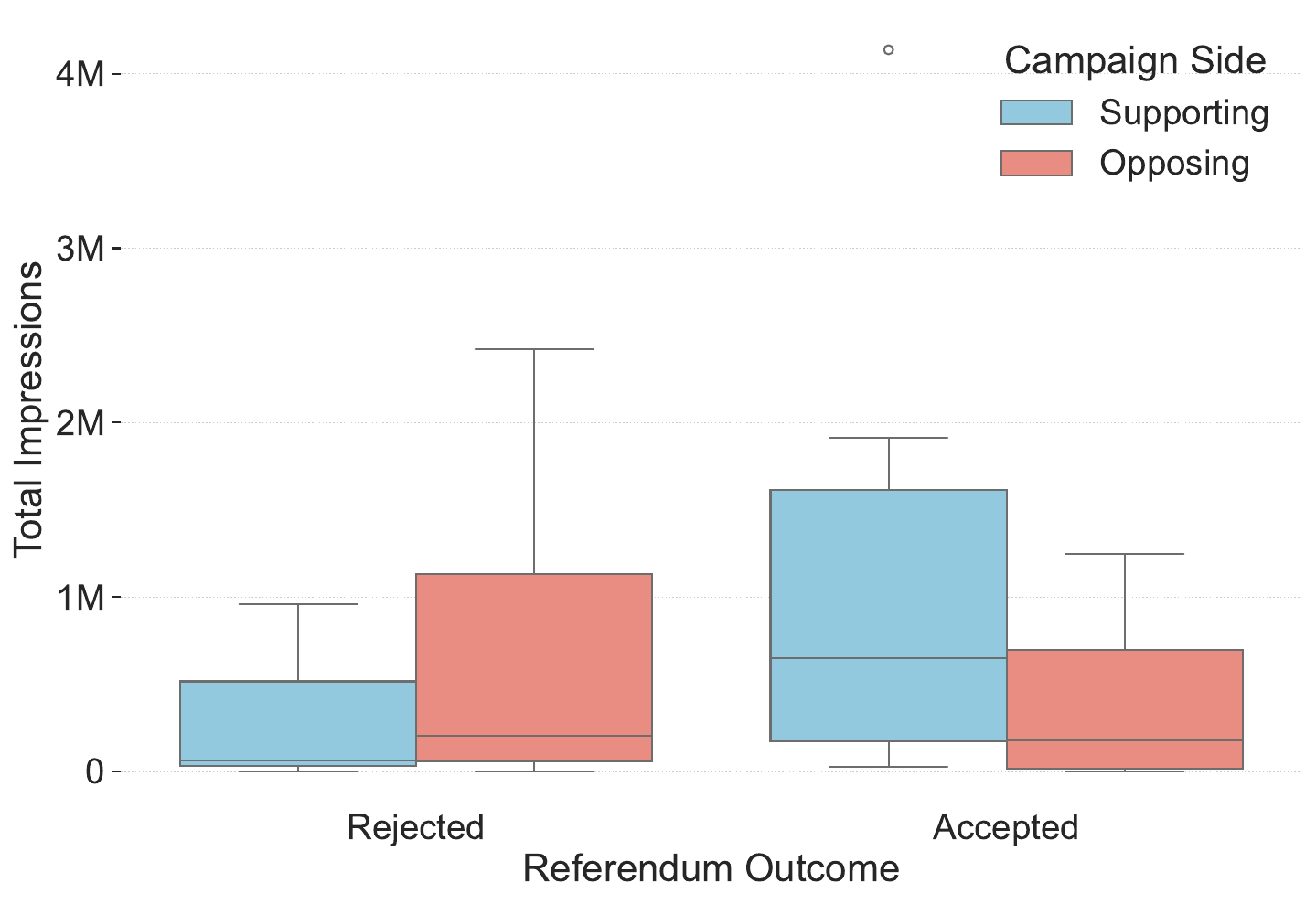}
  \caption{\textbf{Distribution of ad impressions for pro-Yes and pro-No campaigns, grouped by referendum outcome.}}
  \label{fig:outcome_vs_impressions}
\end{figure}

These findings lead to the next research question (\textbf{RQ2.b}): Given the increase in ad activity before referenda, is it possible to explore and measure the influence of referendum ads on the outcome of the vote?
Figure~\ref{fig:outcome_vs_impressions} shows the distribution of all ad impressions supporting or opposing the referenda in relation to their outcomes. This figure suggests that there is a relationship between the number of impressions per side and referendum outcomes. When a referendum is rejected, the number of views of ads opposing the referendum looks to be higher. In contrast, if the outcome of a referendum is positive, the number of views of ads supporting it tends to be higher. 
Correlation analysis shows a moderate positive association between the ratio of Yes to No ad impressions and referendum approval outcome (Spearman $\rho = 0.39$, $p = 0.043$). This indicates that referendums with relatively more Yes-side advertising tended to be approved more often. Then, a Mann-Whitney U non-parametric test is used to independently compare the ``Yes'' and ``No'' impression volumes. The results suggest that approved referenda have a significantly higher volume of ``Yes'' impressions compared to rejected referenda ($U = 146.0$, $p = 0.0135$). In contrast, the volume of ``No'' impressions is not a statistically significant predictor of a referendum rejection ($U = 117.0$, $p = 0.1906$).
In other words, greater exposure to pro-Yes advertising is associated with a higher probability of referendum approval, whereas exposure to pro-No advertising does not show a systematic link to rejection outcomes.
It is important to note that this is a correlation and does not prove causation.
In addition, popularity can act as a confounding factor: popular referendums are more likely to attract higher ad spending for the ``Yes'' side and are also more likely to pass.
This is because ``Yes'' campaigns inherently face an uphill battle against the well-documented ``status quo bias'', in which uncertain voters tend to vote ``No'' to avoid uncertainty~\cite{Bowler1998}. A ``Yes'' campaign must first have strong latent public support to overcome this bias and succeed. Thus, ad volume may indicate a well-funded campaign with this support rather than being an independent driver of the outcome~\cite{Gerber1998}.

\begin{figure*}[ht]
  \centering
  \includegraphics[width=.99\textwidth]{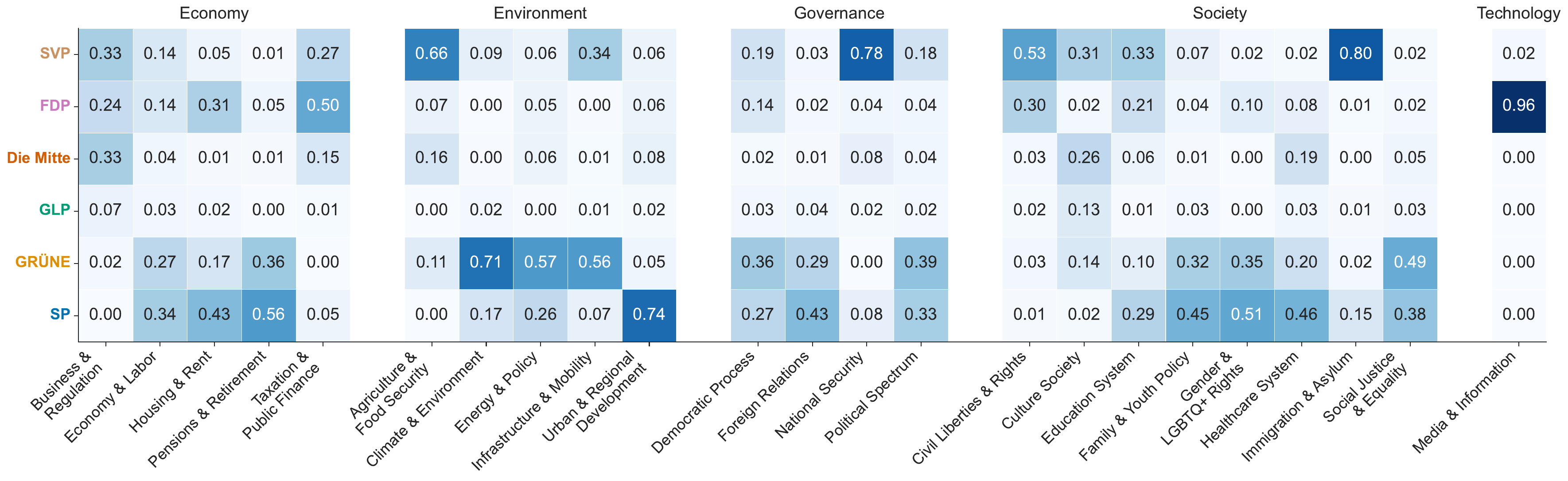}
  \caption{\textbf{Share of impressions per topic by party.} Only the top ten topics for each party are represented.}
  \label{fig:top_topic_correlation}
\end{figure*}

\subsection{\textbf{Federal elections}}

The third research question focuses on the federal elections in October 2023 and the issues promoted in ad campaigns. 
As described in the Federal Election Subsection, each ad published in the month before the federal elections is labeled with a relevance flag and with up to 3 keywords related to the topics discussed in the ad.
Thanks to this data-processing procedure, it is possible to investigate whether parties discuss different issues during federal elections or promote different views on the same issues (\textbf{RQ3.a}).

Figure~\ref{fig:top_topic_correlation} shows the share of impressions by party for the top ten topics by impressions.
Most issues are promoted exclusively by one party or by a few parties that are close to each other on the political spectrum.
For example, the issues of ``Business \& Regulation'' and ``Taxation \& Public Finance'' are almost only represented by the SVP, the FDP, and Die Mitte.
Most of the impressions on the issue of ``Civil Liberties \& Rights'' (which includes topics such as abortion, drugs liberalization, and freedom of speech) are shared by the SVP and the FDP; on the other hand, most of the impressions on the issue ``Agriculture \& Food Security'' are shared by the SVP and Die Mitte.
Instead, issues related to ``Pensions \& Retirement'', ``Climate \& Environment'', ``Energy \& Policy'', and ``Foreign Relations'' (which covers topics such as EU relations, NATO, or the war in Ukraine) are shared almost exclusively by the SP and the GRÜNE. They also focus heavily on ``Family \& Youth Policy'', ``Gender \& LGBTQ+ Rights'', ``Social Justice \& Equality'', and ``Political Spectrum'' (including topics such as progressive policies and liberalism). However, other issues transcend traditional political divides, receiving support from parties that would otherwise be considered ideologically distant. These include ``Education System'' (shared by the SVP, the FDP, and the SP), ``Infrastructure \& Mobility'' (shared by the GRÜNE and the SVP), Housing \& Rent'' (shared by the SP, the GRÜNE and the FDP), ``Healthcare System'' (shared by the SP, the GRÜNE and Die Mitte), and ``Economy \& labour'' (shared by all parties except the GLP and Die Mitte).

Finally, some issues are pushed by a single party.
For instance, ``National Security'' and ``Immigration \& Asylum'' are cornerstones of the SVP's agenda but are not promoted by other parties.
The issue of ``Media \& Information'' (which includes topics such as cancel culture, disinformation, and woke culture) is almost exclusively promoted by the center-right FDP party.
The issue of ``Urban \& Regional Development'' is addressed almost exclusively by the SP.
It is also worth mentioning that the more centrist parties, GLP and Die Mitte, have the lowest share values. It is reasonable to hypothesize that this is because they address a wide range of issues instead of focusing on a few specific topics.
In summary, of the two hypotheses presented at the beginning of this subsection, the most accurate is that parties tend to have different agendas and rarely promote or debate the same topics.

To further investigate similarities and differences between parties, a machine learning model is built to try to predict the party author of an ad (\textbf{RQ3.b}). For this multi-class classification task, a multinomial logistic regression model is trained on the ads relevant to the 2023 federal elections.
For completeness, three different feature combinations are tested: (1) \textit{Demographic}, representing the percentage of impressions per age and gender bucket; (2) \textit{Topic}, which are one-hot encoded topic categories; and (3) a combination of \textit{Demographic + Topics}.

\begin{table}[h!]
  \caption{AUC ROC score for each feature combination/party}
  \label{tab:model_auc}
    \centering
    \begin{tabular}{lccc}
        \toprule
         & Demographic & Topics & Demographic + Topics\\
        \midrule
        SVP & 0.748 & 0.804 & 0.869 \\
        GRÜNE & 0.728 & 0.775 & 0.857 \\
        SP & 0.661 & 0.776 & 0.816 \\
        GLP & 0.653 & 0.736 & 0.795 \\
        FDP & 0.685 & 0.710 & 0.792 \\
        Die Mitte & 0.632 & 0.722 & 0.759 \\
        \bottomrule
    \end{tabular}
\end{table}

In general, the model achieves strong discriminative power across all classes and feature combinations, as shown by Table~\ref{tab:model_auc} where is reported the One-vs-Rest goodness of fit measured as the area under the ROC curve. \textit{Demographic} features alone are sufficient to distinguish the parties with good performance. However, \textit{Topic} features appear to be stronger predictors, and the best-performing model uses the combined set of all features. In the combined model, all parties achieve an AUC ROC score above 0.75, with top-performing classes like SVP and GRÜNE exceeding 0.85.

\begin{figure}[h!]
    \centering
    \begin{minipage}{0.4\linewidth}
        \centering
        \includegraphics[width=\linewidth]{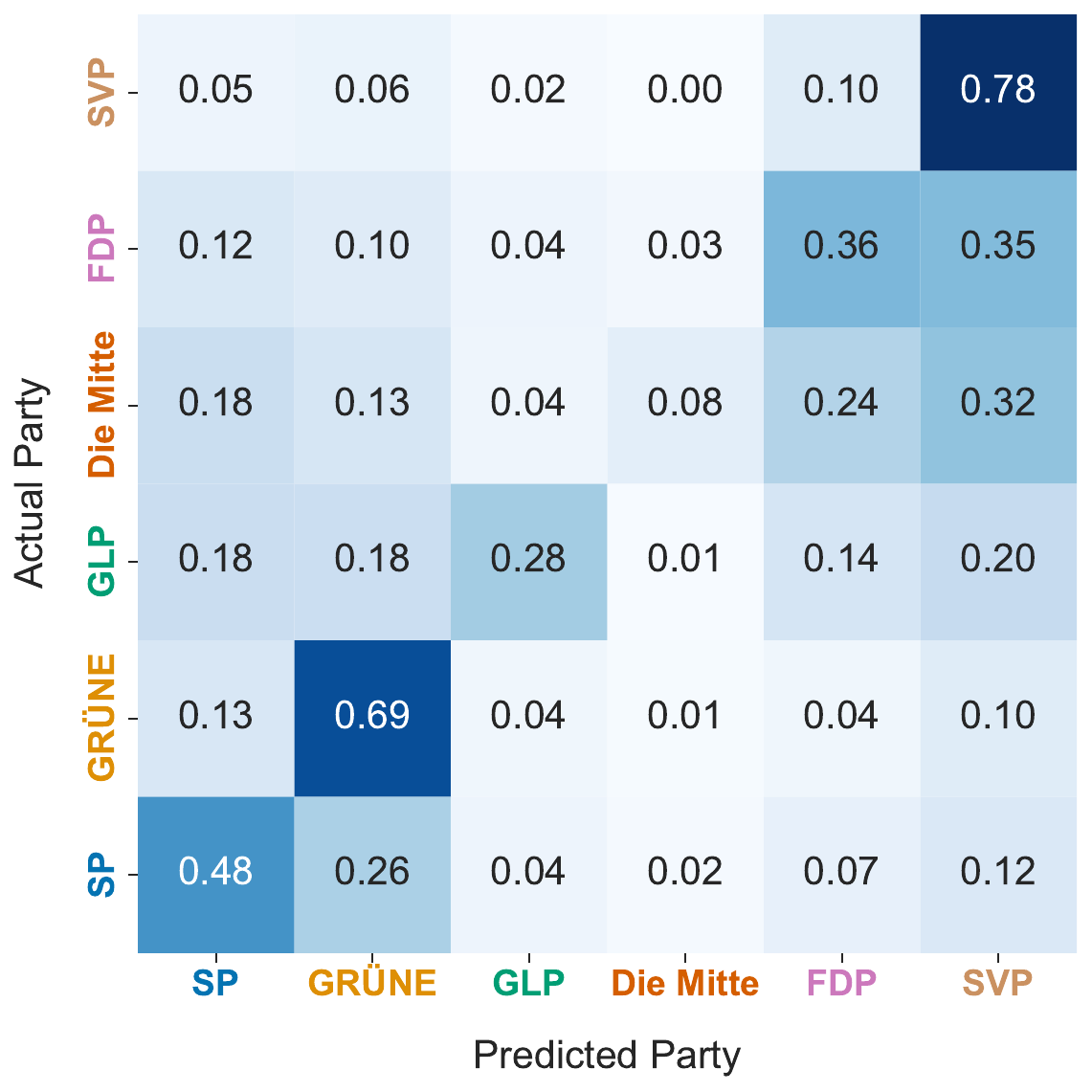}
        \vspace{0.2em} \\
        \textbf{(a)}
        \label{fig:recall_matrix}
    \end{minipage}
    \hfill 
    \begin{minipage}{0.4\linewidth}
        \centering
        \includegraphics[width=\linewidth]{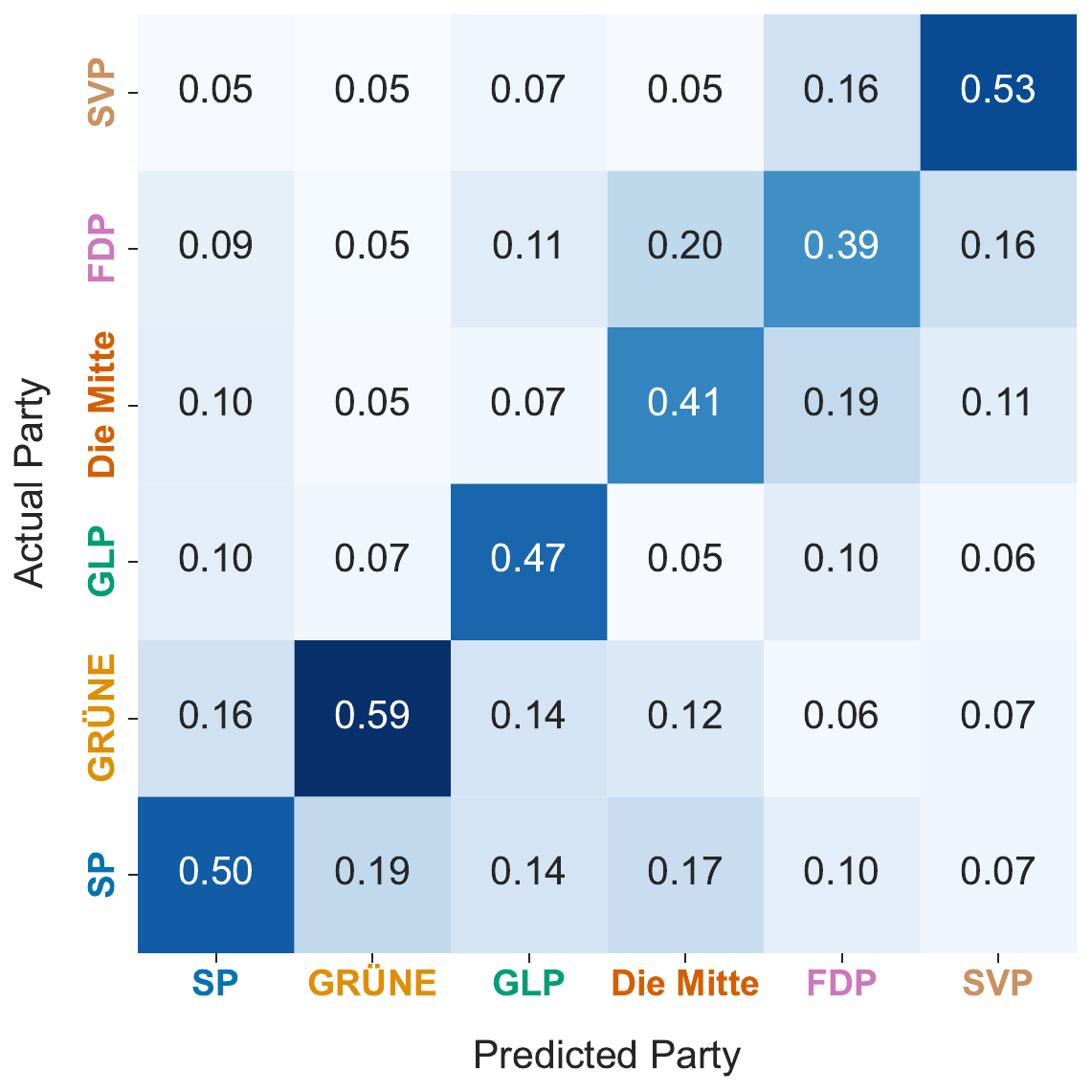}
        \vspace{0.2em} \\
        \textbf{(b)}
        \label{fig:precision_matrix}
    \end{minipage}
    
    \caption{\textbf{Recall-normalized (a) and Precision-normalized (b) confusion matrices.} The metrics refer to the \textit{Demographic + Topics} model. The diagonal shows the per-class recall (a) or precision (b).}
    \label{fig:confusion_matrices}
\end{figure}

The recall-normalized confusion matrix (Figure~\ref{fig:confusion_matrices}.a) highlights a more complex situation than the model's high discriminative power might suggest.

While recall is high for parties at the poles of the political spectrum, such as the SVP (0.78), it is notably low for the centrist parties Die Mitte and the GLP. This is particularly evident for Die Mitte, which achieves a recall of only 0.08. The confusion matrix shows that Die Mitte ads are most frequently misclassified as SVP (32\%) and FDP (24\%).
\begin{figure*}[h!]
  \centering
  \includegraphics[width=.99\textwidth]{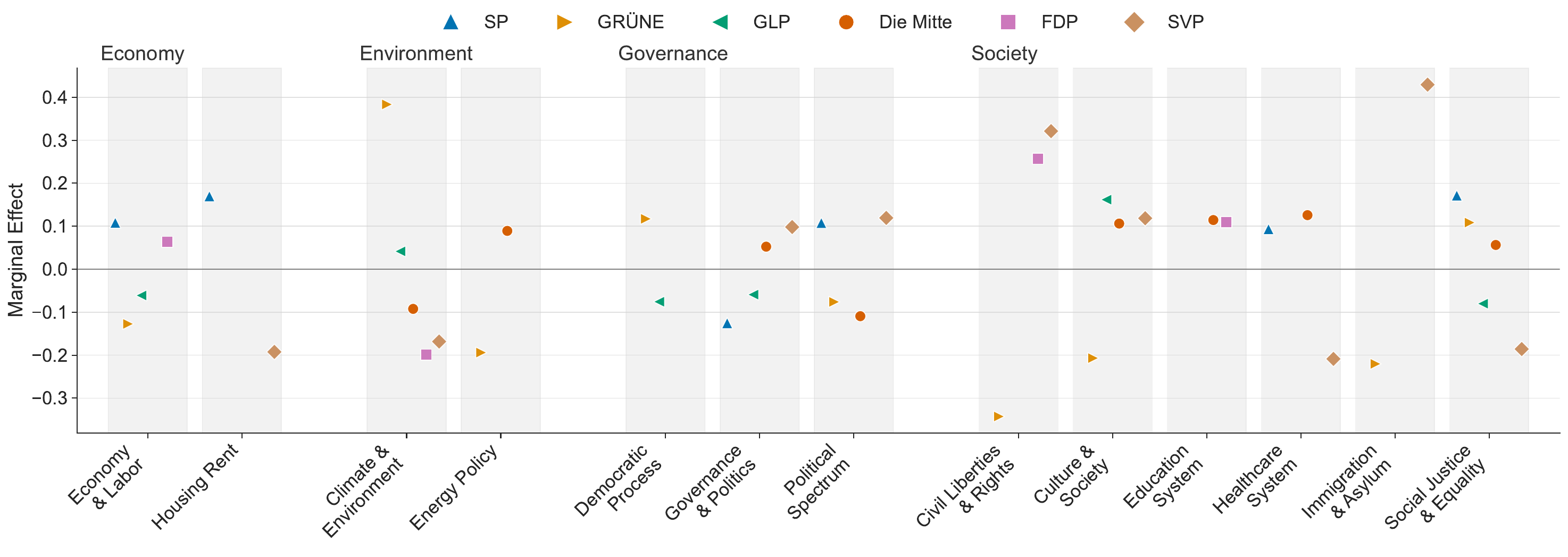}
  \caption{\textbf{Marginal effects of the model with the best feature combination (demographic + topics).} Positive values indicate a higher likelihood of an ad belonging to a specific party when that topic is present. Only statistically significant coefficients ($p < 0.05$) are displayed.}
  \label{fig:model_coefficients} 
\end{figure*}
This result suggests that while the model can reliably distinguish Die Mitte from non-Die Mitte ads (as shown by its high AUC of $0.76$), it fails to classify them as the top choice, probably because of a strong feature overlap with other parties. In line with the analyses conducted for RQ1.a and RQ1.b, the difficulty of predicting centrist parties reinforces the observation that they rely less on microtargeting strategies and avoid focusing on specific issues.

As a final step, the marginal effects of the model trained only on \textit{Topic} features are studied to better understand the predictive interaction between topics and parties.
The topic-share plot in Figure~\ref{fig:top_topic_correlation} represents the raw association between a topic and a party, meaning it also includes the effects of all other topics.
In contrast, marginal effects isolate the predictive effect of adding a single topic while holding all others constant, thereby highlighting the influence of each topic.
Figure~\ref{fig:model_coefficients} presents the marginal effects that are statistically significant ($p < 0.005$), revealing several patterns. For example, environmental and social issues sharply divide the parties. In particular, topics related to ``Climate \& Environment'' are negatively associated with the SVP, the FDP, and Die Mitte, while ``Energy Policy'' topics are a positive predictor only for Die Mitte. In contrast, ``Social Justice \& Equality'' issues are strong positive predictors of the SP and the GRÜNE.
Other policy areas also show distinct alignments. ``Healthcare System'' topics are positively associated with the SP and Die Mitte, and ``Education System'' issues appear as relevant predictive topics for the FDP and Die Mitte. Moreover, ``Immigration \& Asylum'' issues are a significant, strong, positive predictor exclusively for the right-wing party SVP.

\section{Discussion}

The 2024 European Parliament Youth Survey~\cite{Eurobarometer2024} reports that 42\% of European youth (ages 16–30) rely primarily on social media platforms such as TikTok, Instagram, and YouTube for political and social news. Consequently, the paid advertising and algorithmic mechanisms that govern visibility on these platforms now have a disproportionate influence on the democratic process.
Switzerland's system of direct democracy, characterized by frequent popular referendums, is particularly vulnerable to these shifts in democratic deliberation. It is little wonder that online political advertising has shifted from a peripheral tool to a central mechanism for influencing both elections and policy outcomes.

Against this backdrop, this article provides a comprehensive analysis of 40,000 political advertisements (560 million impressions) published on Meta from 2021 to 2025.
A critical finding is the strong correlation between the volume of ``pro-Yes'' advertising and referendum approval. Unlike elections, where voters rely on party heuristics, referendums often involve complex technical issues in which the availability of information plays a decisive role. The correlation between higher ad exposure and favorable outcomes raises normative concerns. If visibility, purchased through an algorithmic auction, can influence policy decisions, then the ``marketplace of ideas'' could become a ``marketplace of budgets''. Although our data reveals a correlation rather than a direct causal relationship, it suggests that well-funded interest groups may use algorithmic reach to circumvent traditional deliberation, especially in ``low-salience'' referendums where voters have fewer preconceived notions.
Furthermore, evidence of demographic microtargeting confirms that political parties are actively segmenting the electorate, rather than merely broadcasting messages. Their adaptation of strategies to Switzerland’s linguistic and regional divisions demonstrates the sophistication of these campaigns. As previous studies have shown, Swiss heterogeneity is a primary driver of political divisions~\cite{OeschRennwald, Mantegazzi03072021} and voting behavior~\cite{SteenbergenMarco, CamatarriFavero}. Our results suggest that algorithms are effectively automating the exploitation of these societal fault lines. By delivering different narratives to young urban women than to older rural men, microtargeting reinforces existing socioeconomic and cultural divisions, which could lead to the polarization seen in other multicultural democracies.

The content analysis reveals a distinct pattern of divergent issues, where political actors ``talk past each other''. Rather than engaging in a shared debate, parties retreat into safe topics, effectively creating parallel public spheres. This behavior aligns with the theory of issue ownership~\cite{Petrocik}, in which parties emphasize topics on which they have a reputational advantage~\cite{BudgeFarlie1983}.
By selectively focusing on ``their'' issues, parties avoid competing on unfavorable ground and instead attempt to make their own issue priorities the most important for voters~\cite{amoros2013issue, BELANGER2008477}.
Social media algorithms likely amplify this effect beyond what is possible in traditional media. In a television broadcast, viewers are exposed to the opposition's arguments, even if they disagree with them. In the algorithmic environment observed, parties can target their ``owned'' issues exclusively to sympathetic demographics, insulating their base from counterarguments. This suggests that digital microtargeting transforms issue ownership from a campaign tactic into a mechanism of epistemic closure.

These conclusions should be interpreted in light of the limitations imposed by the Meta Ad Library API. The current transparency tools only reveal basic targeting parameters, such as age, gender, and location, and omit the vast array of behavioral and interest-based attributes used for profiling~\cite{ZagheniGummadi, MejovaWeberBenevenuto}. Additionally, the data cannot distinguish between targeting decisions made by campaign managers and those optimized automatically by Meta’s delivery algorithms. This ``black box'' is a significant barrier to accountability. Without access to the precise logic of audience selection, researchers cannot determine whether elections are being manipulated by malicious actors or merely optimized by profit-driven algorithms.

This study contributes to the ongoing debate on the governance of the digital public sphere by providing a novel, data-driven analysis of the impact of political advertising on social media on direct democracy.
Although censoring political ads entirely may be undesirable, maintaining the current level of secrecy is not an option. Empowering users and researchers requires shifting from voluntary transparency to comprehensive data access. Citizens can only move from being passive recipients of tailored messaging to active participants in a shared democratic process by understanding how they are being targeted.

\section{Materials and methods}

This study relies on data from two main sources: the Swiss Federal Statistical Office\footnote{\url{https://abstimmungen.admin.ch/en/overview}} and the Meta Ad Library API\footnote{\url{https://www.facebook.com/ads/library/api/}}. The Federal Statistical Office provides detailed statistics related to referenda, including results by canton, participation percentages, and full descriptions of referendum questions. 

\subsection{\textbf{Meta Ad Library}}

The Meta Ad Library API discloses advertisements about social issues, election, or politics that were delivered on Facebook or Instagram during the past seven years. Although the API allows for specifying search parameters, such as limiting results to a specific country, all queries must contain keywords or author names.
The following information can be extracted for each advertisement: the publication date; the end time of the campaign; the textual content; and the number of impressions (provided as an interval). The term ``impressions'' refers to non-unique visualizations. It is also possible to access the campaign expenditure (again presented as an interval) and the audience's demographic profile, specifically their gender (male, female, or unknown) and age (categorized into seven buckets). In line with existing literature~\cite{capozzi2021clandestino}, the midpoint of the interval's boundaries is computed for values presented as ranges (expenditure and impressions). For unbounded ranges, only the defined boundary is used.
To quantify the extent of demographic microtargeting, two statistical measures are employed. First, gender disparities are analyzed using the odds ratio (OR). This metric measures the likelihood of an advertisement being shown to a male user versus a female user. An OR of $1.0$ indicates a perfectly balanced gender distribution; values greater than $1.0$ indicate a bias toward a male audience, while values less than $1.0$ indicate a bias toward a female audience.
Second, differences in age targeting are evaluated using the Wasserstein distance (WD), often referred to as the Earth Mover's Distance. Unlike comparing simple averages, the WD accounts for the entire shape of the probability distributions. It quantifies the minimum amount of ``work'' required to transform one distribution into another. In the context of this study, the WD effectively represents the mean absolute difference in years between the age profile of a specific party's audience and that of the comparison group (e.g., the rest of the dataset).

The collection is initiated using a broad set of election-related keywords (see Table~\ref{tab:keywords}) in German, French, and Italian due to the API's requirement for keyword- or author-based searches. This process yielded an initial corpus of 40,044 ads (22,494 from German keywords, 10,794 from French, and 6,756 from Italian) published between January 1, 2021, and September 30, 2025.
The API's country filter includes advertisements displayed in Switzerland, even if they are intended for a different audience. To create a Swiss-centric dataset, ads are further filtered to include only those for which more than 50\% of impressions come from users based in Switzerland.
This reduces the dataset to 34,559 ads from 2,246 unique pages, which received an estimated 560 million impressions and cost over CHF 4.5 million.

Although the API provides a page name (username) for each ad, this identifier is frequently ambiguous or does not directly correspond to a particular political party, even when the content of the page clearly promotes a party's agenda.
As shown in Figure~\ref{fig:timeseries_impressions}, the 2023 federal elections represent the most significant activity peak in the dataset.
For this reason, the 1,224 pages that published at least one ad in the 30 days preceding the federal election are manually labeled to establish whether they are related to a specific party. This process identified 943 pages that could be directly attributed to a political party, while the remaining 281 pages belong to entities such as media organizations (newspapers and daily newspapers), NGOs, neighborhood committees, and non-partisan referendum groups.
Of the 943 political pages, 17 major parties and smaller entities (e.g., municipal parties and civic lists) are identified. However, only six parties --- SVP, SP, FDP, Die Mitte, GRÜNE, and GLP --- play a dominant role in the advertising landscape, accounting for 95\% of all impressions. Collectively, these parties also secured 91.84\% of the votes in the 2023 federal elections. Tables~\ref{tab:impressions},~\ref{tab:number_ads}, and~\ref{tab:expenditure_ads} provide detailed statistics on impressions, volume, and ad expenditure broken down by party and language.

\subsection{\textbf{Content Annotation}}

\label{Content Annotation}
The filtered dataset of 34,559 ads is cleaned to remove entries with blank or duplicate textual content, resulting in 13,324 unique ads. The paper presents two analyses of the textual content: one focused on referenda, and the other on the 2023 federal elections.

\subsubsection{Referenda}
\label{CA_Referenda}
The first step is to identify which advertisements relate to one of the 42 referenda held between January 2021 and October 2025. The process begins by filtering the ads to include only those containing at least 10 words and published within 30 days of the referendum. This procedure results in 9,118 ads.
The second step involves prompting GPT-4o with the text of the ad campaign to establish whether it is related to a referendum and, if so, what stance the campaign is taking. In addition, the model is provided with the official title and description of each referendum, as published by the Federal Statistical Office. The model checkpoint is ``gpt-4o-2024-08-06'', and the API queries are executed on October 8, 2025. The task is structured as a few-shot query, and the output for each ad is a Boolean parameter indicating whether the ad is relevant to the referendum. If an ad is deemed relevant, the output includes also the referendum's title and the ad's stance (Yes, No, or Neutral). Manual label verification is performed on a random sample of 100 ads, and this verification revealed an accuracy rate of 100\%.

To evaluate the intensity of advertising activity, a Wilcoxon signed-rank test is employed to determine if there is a significant increase in the number of ads and expenditure in the period preceding a referendum. The Wilcoxon signed-rank test is a non-parametric statistical hypothesis test used to compare two related samples, matched samples, or repeated measurements on a single sample to assess whether their population mean ranks differ. In this study, it is used to compare the 30-day ``pre-referendum'' window against the 30-day ``post-referendum'' window for each vote. This test is particularly suitable because it does not assume a normal distribution of the data, which is necessary given the highly skewed nature of advertising impressions and expenditure.

However, the Wilcoxon test only identifies a difference between two time blocks and does not account for external confounding factors such as seasonality, day-of-the-week effects, or long-term growth trends in the dataset. To isolate the specific impact of referendum proximity, a Generalized Additive Model (GAM) is deployed. A GAM is a flexible regression technique that extends Generalized Linear Models by allowing the linear predictor to depend on unknown smooth functions of some predictor variables. By using smoothers for the ``days until referendum'', the day of the year, and the general trend, the model can separate the temporary spike caused by a political campaign from recurring temporal patterns.

\subsubsection{Federal Election}
\label{CA_Federal Election}

The second analysis focuses on the 2023 federal elections. Unlike a referendum, which addresses one or a few political issues, federal elections involve political parties promoting broader agendas. Therefore, it is necessary to not only identify election-relevant ads but also determine the specific topic(s) they cover.
Consistent with the previous task, only ads with at least 10 words published within 30 days before the federal election are selected. The model checkpoint and the API query date are the same of the previous task. The model's output includes a Boolean parameter identifying whether the ad is relevant to the federal election, as well as up to three English keywords related to the ad's topic.
This initial extraction process yielded a large set of 739 unique keywords, which require consolidation into fewer categories.
To this end, each keyword is encoded into a semantic vector using OpenAI's state-of-the-art text-embedding-3-large model. These vectors are then clustered into 25 topics using the K-Means clustering algorithm~\cite{mcqueen1967some}.
This unsupervised method partitions the semantic vectors into $K=25$ distinct clusters by iteratively assigning each data point to the nearest centroid, thereby minimizing the within-cluster variance.
To ensure reliability, qualitative validation is performed on a random subset of 100 ads. In this review, human annotation agreed with model-generated keywords in 96\% of the cases.

To address \textbf{RQ3.b}, a multinomial logistic regression model is employed to predict the political party responsible for a given ad. This method generalizes binary logistic regression to multi-class classification problems by using the softmax function to model the probability distribution over $K$ possible outcomes (parties). For a given input vector $\mathbf{x}$ (containing demographic or topic features), the probability that the ad belongs to party $k$ is given by:
\begin{equation}
    P(y=k \mid \mathbf{x}) = \frac{e^{\mathbf{w}_k^T \mathbf{x}}}{\sum_{j=1}^{K} e^{\mathbf{w}_j^T \mathbf{x}}}
\end{equation}
where $\mathbf{w}_k$ is the weight vector associated with class $k$.
Ads are weighted according to their total number of impressions to ensure that ads with a larger reach have a proportionally larger impact on the model’s parameters.

A multinomial logistic regression model provides interpretable coefficients that reveal the relationship between specific features (e.g., a specific topic) and the likelihood of an ad originating from a specific party. An ablation study is conducted to evaluate feature importance using three configurations: only demographic features, only topic features, and a combination of the two.

The model's performance is evaluated using the Area Under the Receiver Operating Characteristic curve (AUC ROC). The ROC curve plots the true positive rate (sensitivity) against the false positive rate (1-specificity) at various threshold settings. The AUC represents the probability that the model ranks a randomly chosen positive instance higher than a randomly chosen negative instance. Since this is a multi-class problem, the AUC is calculated using a One-vs-Rest strategy, where the performance for each party is evaluated against all other parties combined. An AUC score of 0.5 corresponds to random guessing, indicating a lack of discriminative ability. In contrast, a score of 1.0 signifies perfect classification.
To better understand the model's specific strengths and weaknesses, the recall and precision confusion matrices are also analyzed. Recall quantifies the model's ability to correctly identify all ads belonging to a specific party (True Positive Rate). Precision, on the other hand, measures the accuracy of the model's positive predictions for a given party (Positive Predictive Value).

A multinomial logistic regression model provides interpretable coefficients that reveal the relationship between specific features (e.g., a specific topic) and the likelihood of an ad originating from a specific party. An ablation study is conducted to evaluate feature importance using three configurations: only demographic features, only topic features, and a combination of the two.

The model's performance is evaluated using the Area Under the Receiver Operating Characteristic curve (AUC ROC). The ROC curve plots the true positive rate (sensitivity) against the false positive rate (1-specificity) at various threshold settings. The AUC represents the probability that the model ranks a randomly chosen positive instance higher than a randomly chosen negative instance. Since this is a multi-class problem, the AUC is calculated using a One-vs-Rest strategy, where the performance for each party is evaluated against all other parties combined. An AUC score of 0.5 corresponds to random guessing, indicating a lack of discriminative ability. In contrast, a score of 1.0 signifies perfect classification.
To better understand the model's specific strengths and weaknesses, the recall and precision confusion matrices are also analyzed. Recall quantifies the model's ability to correctly identify all ads belonging to a specific party (True Positive Rate). Precision, on the other hand, measures the accuracy of the model's positive predictions for a given party (Positive Predictive Value).


\section{Acknowledgments}
I thank Dr. Dirk Helbing for fruitful discussions.

\section*{Data Availability}
Data and code are available in the Zenodo repository at \url{https://doi.org/10.5281/zenodo.18256330}.

\section{Supplementary Material}

\begin{table*}[h!]
    \centering
    \caption{Thematic Keyword Groups for Data Collection}
    \label{tab:keywords}
    \begin{tabular}{llll}
    \toprule
    \textbf{English Theme} & \textbf{German Keywords} & \textbf{Italian Keywords} & \textbf{French Keywords} \\
    \midrule
    To vote / To elect & ``wählen'', ``wähle'', ``wählen wir'' & ``votare'', ``vota'', ``votiamo'', & ``voter'', ``vote'', ``votons'', \\
     & & ``eleggere'', ``eleggi'', ``eleggiamo'' & ``élire'', ``élis'', ``élisons'' \\
    \addlinespace 
    Election(s) / Vote(s) & ``Wahl'', ``Wahlen'' & ``elezione'', ``elezioni'', & ``élection'', ``élections'', \\
     & & ``votazioni'', ``voto'' & ``votes'', ``vote'' \\
    \addlinespace
    Poll / Ballot & ``Abstimmungen'', ``Abstimmung'' & \textit{(n/a)} & \textit{(n/a)} \\
    \addlinespace
    Referendum & ``Referendum'' & ``referendum'' & ``référendum'' \\
    \addlinespace
    Electoral Campaign & ``Wahlkampf'', ``Wahlkampagne'' & ``campagna elettorale'' & ``campagne électorale'' \\
    \addlinespace
    Electoral Round & ``Wahlgang'' & ``turno elettorale'' & ``tour électoral'' \\
    \addlinespace
    Counting / Tally & ``Auszählung'' & ``scrutinio'' & ``scrutin'' \\
    \bottomrule
    \end{tabular}
\end{table*}

\begin{table}[h!]
  \caption{Impressions by party and language}
  \label{tab:impressions}
  \centering
        \begin{tabular}{lllllll}
        \toprule
         & SP & GRÜNE & GLP & Die Mitte & FDP & SVP \\
        \midrule
        All & 90.0M & 55.9M & 20.3M & 20.3M & 49.9M & 45.4M \\
        DE & 59.4M & 37.1M & 13.8M & 14.7M & 27.6M & 34.8M \\
        FR & 18.0M & 15.6M & 5.5M & 3.8M & 19.2M & 6.6M \\
        IT & 12.6M & 3.2M & 1.0M & 1.8M & 3.1M & 4.0M \\
        \bottomrule
    \end{tabular}
\end{table}

\begin{table}[h!]
  \caption{Number of ads by party and language}
  \label{tab:number_ads}
  \centering
    \begin{tabular}{lllllll}
        \toprule
         & SP & GRÜNE & GLP & Die Mitte & FDP & SVP \\
        \midrule
        All & 3.2K & 2.6K & 1.4K & 1.4K & 3.0K & 2.5K \\
        DE & 2.1K & 1.7K & 1.1K & 1.0K & 2.0K & 2.1K \\
        FR & 0.7K & 0.7K & 0.2K & 0.3K & 0.9K & 0.2K \\
        IT & 0.4K & 0.1K & 0.1K & 0.1K & 0.1K & 0.2K \\
        \bottomrule
    \end{tabular}
\end{table}

\begin{table}[h!]
  \caption{Expenditure on ads by party and language}
  \label{tab:expenditure_ads}
  \centering
    \begin{tabular}{lllllll}
        \toprule
         & SP & GRÜNE & GLP & Die Mitte & FDP & SVP \\
        \midrule
        All & 758.4K & 353.1K & 157.6K & 188.5K & 372.7K & 353.0K \\
        DE & 484.2K & 229.1K & 113.9K & 138.7K & 226.1K & 286.9K \\
        FR & 149.5K & 103.5K & 37.9K & 39.4K & 132.2K & 39.4K \\
        IT & 124.7K & 20.5K & 5.8K & 10.3K & 14.4K & 26.6K \\
        \bottomrule
    \end{tabular}
\end{table}

\begin{table}[h!]
  \caption{Wilcoxon signed-rank test of ad views 30 days pre- and post-referendum ($p < 0.001$ for all tests).}
    \label{tab:wilcoxon}
    \centering
    \begin{tabular}{lcc}
        \toprule
         &Impressions & Expenditure \\
        \midrule
        All & 69456 & 68337\\
        DE & 62227 & 64002\\
        FR & 61751 & 58278\\
        IT & 35867 & 33361\\
        \bottomrule
    \end{tabular}
\end{table}

\bibliographystyle{unsrt}  
\bibliography{references}

@inproceedings{mcqueen1967some,
  title={Some methods of classification and analysis of multivariate observations},
  author={McQueen, James B},
  booktitle={Proc. of 5th Berkeley Symposium on Math. Stat. and Prob.},
  pages={281--297},
  year={1967}
}

@inproceedings{capozzi2021clandestino,
    author = {Capozzi, Arthur and De Francisci Morales, Gianmarco and Mejova, Yelena and Monti, Corrado and Panisson, Andr\'{e} and Paolotti, Daniela},
    title = {Clandestino or Rifugiato?Anti-immigration Facebook Ad Targeting in Italy},
    year = {2021},
    isbn = {9781450380966},
    publisher = {Association for Computing Machinery},
    address = {New York, NY, USA},
    doi = {10.1145/3411764.3445082},
    booktitle = {Proceedings of the 2021 CHI Conference on Human Factors in Computing Systems},
    articleno = {179},
    numpages = {15},
    location = {Yokohama, Japan},
    series = {CHI '21}
}

@book{Bowler1998,
  author    = {Bowler, Shaun and Donovan, Todd},
  title     = {Demanding Choices: Opinion, Voting, and Direct Democracy},
  publisher = {University of Michigan Press},
  year      = {1998}
}

@incollection{Gerber1998,
  author    = {Gerber, Elisabeth R.},
  title     = {Pressuring Legislatures through the Use of Initiatives: Two Forms of Indirect Influence},
  booktitle = {Citizens as Legislators: Direct Democracy in the United States},
  editor    = {Bowler, Shaun and Donovan, Todd and Tolbert, Caroline J.},
  publisher = {Ohio State University Press},
  year      = {1998}
}

@article{bond201261,
        title={A 61-million-person experiment in social influence and political mobilization},
        author={Bond, Robert M and Fariss, Christopher J and Jones, Jason J and Kramer, Adam DI and Marlow, Cameron and Settle, Jaime E and Fowler, James H},
        journal={Nature},
        volume={489},
        number={7415},
        pages={295--298},
        year={2012},
        publisher={Nature Publishing Group UK London},
        doi={10.1038/nature11421}
}

@inproceedings{capozzi2020facebook,
  title={Facebook ads: politics of migration in Italy},
  author={Capozzi, Arthur and de Francisci Morales, Gianmarco and Mejova, Yelena and Monti, Corrado and Panisson, Andr{\'e} and Paolotti, Daniela},
  booktitle={International Conference on Social Informatics},
  pages={43--57},
  year={2020},
  organization={Springer},
  doi={10.1007/978-3-030-60975-7_4}
}

@article{Mejova,
        author = {Mejova, Yelena and Capozzi, Arthur and Monti, Corrado and De Francisci Morales, Gianmarco},
        title = {Narratives of War: Ukrainian Memetic Warfare on Twitter},
        year = {2025},
        publisher = {Association for Computing Machinery},
        address = {New York, NY, USA},
        volume = {9},
        number = {2},
        doi = {10.1145/3711037},
        journal = {Proc. ACM Hum.-Comput. Interact.},
        articleno = {CSCW139},
        numpages = {28}
}

@inproceedings{Papakyriakopoulos4,
        author = {Papakyriakopoulos, Orestis and Engelmann, Severin and Winecoff, Amy},
        title = {Upvotes? Downvotes? No Votes? Understanding the relationship between reaction mechanisms and political discourse on Reddit},
        year = {2023},
        isbn = {9781450394215},
        publisher = {Association for Computing Machinery},
        address = {New York, NY, USA},
        doi = {10.1145/3544548.3580644},
        booktitle = {Proceedings of the 2023 CHI Conference on Human Factors in Computing Systems},
        articleno = {549},
        numpages = {28},
        location = {Hamburg, Germany},
        series = {CHI '23}
}

@inproceedings{LeeYu-Hao,
        author = {Lee, Yu-Hao and Hsieh, Gary},
        title = {Does slacktivism hurt activism? the effects of moral balancing and consistency in online activism},
        year = {2013},
        isbn = {9781450318990},
        publisher = {Association for Computing Machinery},
        address = {New York, NY, USA},
        doi = {10.1145/2470654.2470770},
        booktitle = {Proceedings of the SIGCHI Conference on Human Factors in Computing Systems},
        pages = {811–820},
        numpages = {10},
        location = {Paris, France},
        series = {CHI '13}
}

@article{WilliamsNagel,
        author = {Kelly Williams Nagel},
        title = {Make America Meme Again: The Rhetoric of the Alt-Right},
        journal = {Quarterly Journal of Speech},
        volume = {106},
        number = {2},
        pages = {216--219},
        year = {2020},
        publisher = {NCA Website},
        doi = {10.1080/00335630.2020.1744818}
}

@article{Leon,
        title={The Rise of Meme Culture: Internet Political Memes as Tools for Analysing Philippine Propaganda},
        volume={2},
        url={https://jcsll.gta.org.uk/index.php/home/article/view/70},
        DOI={10.46809/jcsll.v2i4.70},
        number={4},
        journal={Journal of Critical Studies in Language and Literature},
        author={G. De Leon ,
        Faye Margarette and Ballesteros-Lintao ,
        Rachelle},
        year={2021},
        month={May},
        pages={1-13} 
}

@incollection{ANDERSON201541,
        title = {Chapter 2 - The Advertising-Financed Business Model in Two-Sided Media Markets},
        editor = {Simon P. Anderson and Joel Waldfogel and David Strömberg},
        series = {Handbook of Media Economics},
        publisher = {North-Holland},
        volume = {1},
        pages = {41-90},
        year = {2015},
        booktitle = {Handbook of Media Economics},
        issn = {2213-6630},
        doi = {https://doi.org/10.1016/B978-0-444-62721-6.00002-0},
        author = {Simon P. Anderson and Bruno Jullien}
}

@article{Epstein,
        author = {Robert Epstein  and Ronald E. Robertson },
        title = {The search engine manipulation effect (SEME) and its possible impact on the outcomes of elections},
        journal = {Proceedings of the National Academy of Sciences},
        volume = {112},
        number = {33},
        pages = {E4512-E4521},
        year = {2015},
        doi = {10.1073/pnas.1419828112}
}

@inproceedings{RanDezhi,
        author = {Ran, Dezhi and Zheng, Weiqiang and Li, Yunqi and Bian, Kaigui and Zhang, Jie and Deng, Xiaotie},
        title = {Revenue and User Traffic Maximization in Mobile Short-Video Advertising},
        year = {2022},
        isbn = {9781450392136},
        publisher = {International Foundation for Autonomous Agents and Multiagent Systems},
        address = {Richland, SC},
        pages = {1092–1100},
        numpages = {9},
        location = {Virtual Event, New Zealand},
        series = {AAMAS '22}
}

@article{MatiasNathan,
        author = {Matias, J. Nathan and Hounsel, Austin and Feamster, Nick},
        title = {Software-Supported Audits of Decision-Making Systems: Testing Google and Facebook's Political Advertising Policies},
        year = {2022},
        issue_date = {April 2022},
        publisher = {Association for Computing Machinery},
        address = {New York, NY, USA},
        volume = {6},
        number = {CSCW1},
        doi = {10.1145/3512965},
        journal = {Proc. ACM Hum.-Comput. Interact.},
        articleno = {118},
        numpages = {19}
}

@article{PierriGianmarco,
    author = {Bär, Dominik and Pierri, Francesco and De Francisci Morales, Gianmarco and Feuerriegel, Stefan},
    title = {Systematic discrepancies in the delivery of political ads on Facebook and Instagram},
    journal = {PNAS Nexus},
    volume = {3},
    number = {7},
    pages = {pgae247},
    year = {2024},
    month = {06},
    issn = {2752-6542},
    doi = {10.1093/pnasnexus/pgae247},
    url = {https://doi.org/10.1093/pnasnexus/pgae247},
    eprint = {https://academic.oup.com/pnasnexus/article-pdf/3/7/pgae247/58651815/pgae247.pdf},
}

@inproceedings{Sosnovik,
        author = {Sosnovik, Vera and Goga, Oana},
        title = {Understanding the Complexity of Detecting Political Ads},
        year = {2021},
        isbn = {9781450383127},
        publisher = {Association for Computing Machinery},
        address = {New York, NY, USA},
        doi = {10.1145/3442381.3450049},
        pages = {2002–2013},
        numpages = {12},
        location = {Ljubljana, Slovenia},
        series = {WWW '21}
}

@book{KarpfDavid,
    author = {Karpf, David},
    title = {Analytic Activism: Digital Listening and the New Political Strategy},
    publisher = {Oxford University Press},
    year = {2017},
    month = {01},
    isbn = {9780190266127},
    doi = {10.1093/acprof:oso/9780190266127.001.0001}
}

@inproceedings{AliSapiezynski,
        author = {Ali, Muhammad and Sapiezynski, Piotr and Korolova, Aleksandra and Mislove, Alan and Rieke, Aaron},
        title = {Ad Delivery Algorithms: The Hidden Arbiters of Political Messaging},
        year = {2021},
        isbn = {9781450382977},
        publisher = {Association for Computing Machinery},
        address = {New York, NY, USA},
        doi = {10.1145/3437963.3441801},
        booktitle = {Proceedings of the 14th ACM International Conference on Web Search and Data Mining},
        pages = {13–21},
        numpages = {9},
        location = {Virtual Event, Israel},
        series = {WSDM '21}
}

@article{O’Reilly_Strauss_Mazzucato_2024,
        title={Algorithmic attention rents: A theory of digital platform market power},
        volume={6}, DOI={10.1017/dap.2024.1},
        journal={Data \& Policy},
        author={O’Reilly, Tim and Strauss, Ilan and Mazzucato, Mariana},
        year={2024},
        pages={e6}
}

@article{DommettKatharine,
        author = {Dommett, Katharine and Power, Sam},
        title = {The Political Economy of Facebook Advertising: Election Spending, Regulation and Targeting Online},
        journal = {The Political Quarterly},
        volume = {90},
        number = {2},
        pages = {257-265},
        doi = {https://doi.org/10.1111/1467-923X.12687},
        year = {2019}
}

@article{madrigal2017facebook,
        title={What Facebook did to American democracy},
        author={Madrigal, Alexis C},
        journal={The Atlantic},
        year={2017}
}

@article{entous2017russian,
  title={Russian operatives used Facebook ads to exploit America’s racial and religious divisions},
  author={Entous, Adam and Timberg, Craig and Dwoskin, Elizabeth},
  journal={Washington Post},
  volume={25},
  year={2017}
}

@techreport{Eurobarometer2024,
  author      = {{European Parliament}},
  title       = {Youth and Democracy in the European Union: Eurobarometer Survey},
  institution = {European Union},
  year        = {2024},
  month       = {May},
  type        = {Flash Eurobarometer},
  number      = {545},
  url         = {https://europa.eu/eurobarometer/surveys/detail/3216},
  note        = {Accessed: 2025-06-17}
}

@article{BarFeuerriegel,
    author = {Bär, Dominik and Pröllochs, Nicolas and Feuerriegel, Stefan},
    title = {The role of social media ads for election outcomes: Evidence from the 2021 German election},
    journal = {PNAS Nexus},
    volume = {4},
    number = {3},
    pages = {pgaf073},
    year = {2025},
    month = {03},
    issn = {2752-6542},
    doi = {10.1093/pnasnexus/pgaf073},
    url = {https://doi.org/10.1093/pnasnexus/pgaf073},
    eprint = {https://academic.oup.com/pnasnexus/article-pdf/4/3/pgaf073/62519805/pgaf073_supplementary_data.pdf},
}

@InProceedings{EmilioFerrara,
        author="Dutt, Ritam
        and Deb, Ashok
        and Ferrara, Emilio",
        title="``Senator, We Sell Ads'': Analysis of the 2016 Russian Facebook Ads Campaign",
        booktitle="Advances in Data Science",
        year="2019",
        publisher="Springer Singapore",
        address="Singapore",
        pages="151--168",
        isbn="978-981-13-3582-2"
}

@inproceedings{MejovaKalimeri,
        author = {Mejova, Yelena and Kalimeri, Kyriaki},
        title = {COVID-19 on Facebook Ads: Competing Agendas around a Public Health Crisis},
        year = {2020},
        isbn = {9781450371292},
        publisher = {Association for Computing Machinery},
        address = {New York, NY, USA},
        doi = {10.1145/3378393.3402241},
        booktitle = {Proceedings of the 3rd ACM SIGCAS Conference on Computing and Sustainable Societies},
        pages = {22–31},
        numpages = {10},
        location = {Ecuador},
        series = {COMPASS '20}
}

@inproceedings{CapozziThin,
        author = {Capozzi, Arthur and De Francisci Morales, Gianmarco and Mejova, Yelena and Monti, Corrado and Panisson, Andr\'{e}},
        title = {The Thin Ideology of Populist Advertising on Facebook during the 2019 EU Elections},
        year = {2023},
        isbn = {9781450394161},
        publisher = {Association for Computing Machinery},
        address = {New York, NY, USA},
        doi = {10.1145/3543507.3583267},
        booktitle = {Proceedings of the ACM Web Conference 2023},
        pages = {2852–2862},
        numpages = {11},
        location = {Austin, TX, USA},
        series = {WWW '23}
}

@article{CanoLorena,
        title = {Disinformation in Facebook Ads in the 2019 Spanish General Election Campaigns},
        author = {Cano-Orón, Lorena and Calvo, Dafne and López García, Guillermo and Baviera, Tomás},
        journal = {Media and Communication},
        number = {1},
        pages = {217-228},
        volume = {9},
        year = {2021},
        issn = {2183-2439},
        doi = {https://doi.org/10.17645/mac.v9i1.3335},
}

@article{Disset2010,
	title={Homogeneity, Heterogeneity and Direct Democracy: The Case of Swiss Referenda},
	volume={40},
	DOI={10.1017/S0008423907070138},
	number={2},
	journal={Canadian Journal of Political Science},
	author={Diskin, Abraham and Eschet-Schwarz,	André and Felsenthal, Dan S.},
	year={2007},
	pages={317–342}
}

@incollection{Kriesi2005,
	title={Direct democracy: the Swiss experience},
	author={Kriesi, Hanspeter},
	booktitle={Evaluating Democratic Innovations},
	pages={39--55},
	year={2012},
	publisher={Routledge}
}

@book{Linder2010,
	title={Swiss democracy: Possible solutions to conflict in multicultural societies},
	author={Linder, Wolf and Mueller, Sean},
	year={2021},
	publisher={Springer Nature},
	doi={10.1007/978-3-030-63266-3}
}

@article{Papadopoulos2001,
    author = {Yannis Papadopoulos},
    title = {How does direct democracy matter? The impact of referendum votes on politics and policy‐making},
    journal = {West European Politics},
    volume = {24},
    number = {2},
    pages = {35--58},
    year = {2001},
    publisher = {Routledge},
    doi = {10.1080/01402380108425432}
}

@inbook{Sager_Zollinger_2011,
	series={Routledge Advances in European Politics},
	title={The Swiss political system in comparative perspective},
	ISBN={978-0-415-58051-9},
	DOI={10.4324/9780203829394},
	number={7272},
	booktitle={Switzerland in Europe: Continuity and change in the Swiss political economy},
	publisher={Routledge},
	author={Sager, F and Zollinger, C},
	year={2011},
	pages={27–42},
	collection={Routledge Advances in European Politics}
}

@misc{FSOLang,
	author = {Bundesamt für Statistik (BFS)},
	year         = {2023}, 
	month        = {Feb},
	language     = {EN},
	title        = {Main languages, 1970-2023},
	number       = {34247523},
	howpublished = {Web},
	url          = {https://dam-api.bfs.admin.ch/hub/api/dam/assets/34247523/master}
}

@misc{FSOrelsurvey,
	author = {Bundesamt für Statistik (BFS)},
	year         = {2025},
	month        = {Jan},
	language     = {EN},
	title        = {Evolution of religious landscape},
	number       = {33748799},
	howpublished = {Web},
	url          = {https://dam-api.bfs.admin.ch/hub/api/dam/assets/33748799/master}
}

@misc{FSOGeo,
	author = {Bundesamt für Statistik (BFS)},
	year         = {2024},
	month        = {Mar},
	language     = {EN},
	title        = {Three new agglomerations and 10 more towns than 10 years ago: urbanisation continues in Switzerland},
	address      = {Neuchâtel},
	number       = {30665765},
	howpublished = {Web},
	url          = {https://dam-api.bfs.admin.ch/hub/api/dam/assets/30665765/master}
}

@book{FSO_GDP,
    author = {BFS},
    month = {Mar},
    language = {IT},
    title = {PIL in aumento in tutti i Cantoni nel 2022},
    institution = {Bundesamt für Statistik (BFS)},
    publisher = {Bundesamt für Statistik (BFS)},
    address = {Neuchâtel},
    copyright = {Bundesamt für Statistik (BFS)},
    ISBN = {978-3-303-21059-8},
    pages = {4},
    year = {2024},
    number = {30665780},
    url = {https://dam-api.bfs.admin.ch/hub/api/dam/assets/32627363/master},
    organization = {Bundesamt für Statistik}
}

@misc{FSO_Elections2023,
  author    = {{Swiss Federal Statistical Office (FSO)}},
  title     = {National Council elections 2023: Party strength, number of seats, and voter share},
  year      = {2023},
  publisher = {Swiss Federal Statistical Office (FSO)},
  address   = {Neuchâtel, Switzerland},
  note      = {Data table, identifier: je-e-01.02.01.01}
}

@article{ChuVliegenthart,
        author = {Xiaotong Chu and Rens Vliegenthart and Lukas Otto and Sophie Lecheler and Claes de Vreese and Sanne Kruikemeier},
        title = {Do Online Ads Sway Voters? Understanding the Persuasiveness of Online Political Ads},
        journal = {Political Communication},
        volume = {41},
        number = {2},
        pages = {290--314},
        year = {2024},
        publisher = {Routledge},
        doi = {10.1080/10584609.2023.2276104},
}

@article{warner1995direct,
  title={Direct democracy: The right of the people to make fools of themselves; the use and abuse of initiative and referendum, a local government perspective},
  author={Warner, Daniel M},
  journal={Seattle UL Rev.},
  volume={19},
  pages={47},
  year={1995},
  publisher={HeinOnline}
}

@book{gerber2011populist,
    ISBN = {9780691002675},
    author = {Elisabeth R. Gerber},
    publisher = {Princeton University Press},
    title = {The Populist Paradox: Interest Group Influence and the Promise of Direct Legislation},
    year = {1999}
}

@article{BrunoFrey,
 ISSN = {00028282},
 author = {Bruno S. Frey},
 journal = {The American Economic Review},
 number = {2},
 pages = {338--342},
 publisher = {American Economic Association},
 title = {Direct Democracy: Politico-Economic Lessons from Swiss Experience},
 urldate = {2025-11-04},
 volume = {84},
 year = {1994}
}

@book{Serdült2014,
    author={Serd{\"u}lt, Uwe},
    title={Referendums in Switzerland},
    bookTitle={Referendums Around the World: The Continued Growth of Direct Democracy},
    year={2014},
    publisher={Palgrave Macmillan UK},
    address={London},
    pages={65--121},
    isbn={978-1-137-31470-3},
    doi={10.1057/9781137314703\_4}
}

@article{STUTZERALOISBRUNO,
	author={Stutzer, Alois and Frey, Bruno S},
	title = {Political participation and procedural utility: An empirical study},
	journal = {European Journal of Political Research},
	volume = {45},
	number = {3},
	pages = {391-418},
	doi = {https://doi.org/10.1111/j.1475-6765.2006.00303.x},
	year = {2006}
}

@article{FeldLarsSaviozMarcel,
	author = {Feld, Lars P. and Savioz, Marcel R.},
	title = {Direct Democracy Matters for Economic Performance: An Empirical Investigation},
	journal = {Kyklos},
	volume = {50},
	number = {4},
	pages = {507-538},
	doi = {https://doi.org/10.1111/1467-6435.00028},
	year = {1997}
}

@article{FreyBrunoStutzerAlois2000,
	author = {Frey, Bruno S. and Stutzer, Alois},
	title = {Happiness, Economy and Institutions},
	journal = {The Economic Journal},
	volume = {110},
	number = {466},
	pages = {918-938},
	doi = {https://doi.org/10.1111/1468-0297.00570},
	year = {2000}
}

@article{BLONDELJEAN,
	author = {BLONDEL, JEAN},
	title = {The Politics of Switzerland: Continuity and Change in a Consensus Democracy – By H. Kriesi and A. Trexsel},
	journal = {JCMS: Journal of Common Market Studies},
	volume = {47},
	number = {4},
	pages = {931-932},
	doi = {https://doi.org/10.1111/j.1468-5965.2009.02026\_14.x},
	year = {2009}
}

@book{JamiesonKathleenHall,
    author = {Jamieson, Kathleen Hall},
    title = {Dirty Politics},
    publisher = {Oxford University Press},
    year = {1992},
    month = {09},
    isbn = {9780195078541},
    doi = {10.1093/oso/9780195078541.001.0001}
}

@article{ZuiderveenBorgesiusFrederik,
    title = {Online political microtargeting: Promises and threats for democracy},
    author = {Zuiderveen Borgesius, Frederik J. and Judith Moller and Sanne Kruikemeier and Ronan Fathaigh and Kristina Irion and Tom Dobber and Balazs Bodo and de Vreese, Claes},
    year = {2018},
    month = {fe},
    day = {9},
    doi = {10.18352/ulr.420},
    language = {English},
    volume = {14},
    pages = {82--96},
    journal = {Utrecht Law Review},
    issn = {1871-515X},
    publisher = {Utrecht University Library Open Access Journals},
    number = {1}
}

@article{SimchonEdwards,
    author = {Simchon, Almog and Edwards, Matthew and Lewandowsky, Stephan},
    title = {The persuasive effects of political microtargeting in the age of generative artificial intelligence},
    journal = {PNAS Nexus},
    volume = {3},
    number = {2},
    pages = {pgae035},
    year = {2024},
    month = {01},
    issn = {2752-6542},
    doi = {10.1093/pnasnexus/pgae035}
}

@article{TappinWittenberg,
	author = {Ben M. Tappin  and Chloe Wittenberg  and Luke B. Hewitt  and Adam J. Berinsky  and David G. Rand },
	title = {Quantifying the potential persuasive returns to political microtargeting},
	journal = {Proceedings of the National Academy of Sciences},
	volume = {120},
	number = {25},
	pages = {e2216261120},
	year = {2023},
	doi = {10.1073/pnas.2216261120},
}

@article{ZagheniGummadi,
 ISSN = {00987921, 17284457},
 author = {Emilio Zagheni and Ingmar Weber and Krishna Gummadi},
 journal = {Population and Development Review},
 number = {4},
 pages = {721--734},
 publisher = {[Population Council, Wiley]},
 title = {Leveraging Facebook's Advertising Platform to Monitor Stocks of Migrants},
 urldate = {2025-11-04},
 volume = {43},
 year = {2017}
}

@article{DanielaPerrotta,
    author = {Grow, André and Perrotta, Daniela and Del Fava, Emanuele and Cimentada, Jorge and Rampazzo, Francesco and Gil-Clavel, Sofia and Zagheni, Emilio and Flores, René D. and Ventura, Ilana and Weber, Ingmar},
    title = {Is Facebook's advertising data accurate enough for use in social science research? Insights from a cross-national online survey},
    journal = {Journal of the Royal Statistical Society: Series A (Statistics in Society)},
    volume = {185},
    number = {S2},
    pages = {S343-S363},
    doi = {https://doi.org/10.1111/rssa.12948},
    year = {2022}
}

@inproceedings{MejovaWeberBenevenuto,
    author = {Araujo, Matheus and Mejova, Yelena and Weber, Ingmar and Benevenuto, Fabricio},
    title = {Using Facebook Ads Audiences for Global Lifestyle Disease Surveillance: Promises and Limitations},
    year = {2017},
    isbn = {9781450348966},
    publisher = {Association for Computing Machinery},
    address = {New York, NY, USA},
    doi = {10.1145/3091478.3091513},
    pages = {253–257},
    numpages = {5},
    location = {Troy, New York, USA},
    series = {WebSci '17}
}

@article{OeschRennwald,
	author = {Oesch, Daniel and Rennwald, Line},
	title = {The Class Basis of Switzerland's Cleavage between the New Left and the Populist Right},
	journal = {Swiss Political Science Review},
	volume = {16},
	number = {3},
	pages = {343-371},
	doi = {https://doi.org/10.1002/j.1662-6370.2010.tb00433.x},
	year = {2010}
}

@article{Mantegazzi03072021,
        author = {Daniele Mantegazzi},
        title = {The geography of political ideologies in Switzerland over time},
        journal = {Spatial Economic Analysis},
        volume = {16},
        number = {3},
        pages = {378--396},
        year = {2021},
        publisher = {RSA Website},
        doi = {10.1080/17421772.2020.1860251}
}

@article{CamatarriFavero,
	author = {Camatarri, Stefano and Favero, Adrian and Gallina, Marta and Luartz, Lewis},
	title = {The electoral behaviour of voters with migration backgrounds and natives at the 2015 and 2019 Swiss National Elections. Two completely different stories?},
	journal = {Swiss Political Science Review},
	volume = {28},
	number = {2},
	pages = {319-337},
	doi = {https://doi.org/10.1111/spsr.12508},
	year = {2022}
}

@article{SteenbergenMarco,
	author = {Steenbergen, Marco R.},
	title = {Decomposing the Vote: Individual, Communal, and Cantonal Sources of Voting Behavior in Switzerland},
	journal = {Swiss Political Science Review},
	volume = {16},
	number = {3},
	pages = {403-424},
	doi = {https://doi.org/10.1002/j.1662-6370.2010.tb00435.x},
	year = {2010}
}

@article{Petrocik,
 ISSN = {00925853, 15405907},
 author = {John R. Petrocik},
 journal = {American Journal of Political Science},
 number = {3},
 pages = {825--850},
 publisher = {[Midwest Political Science Association, Wiley]},
 title = {Issue Ownership in Presidential Elections, with a 1980 Case Study},
 urldate = {2025-11-04},
 volume = {40},
 year = {1996}
}

@book{BudgeFarlie1983,
  title={Explaining and predicting elections: Issue effects and party strategies in twenty-three democracies},
  author={Budge, Ian and Farlie, Dennis J},
  year={2025},
  publisher={Taylor \& Francis}
}

@article{amoros2013issue,
    title={Issue convergence or issue divergence in a political campaign?},
    author={Amor{\'o}s, Pablo and Puy, M Socorro},
    journal={Public Choice},
    volume={155},
    number={3},
    pages={355--371},
    year={2013},
    doi={10.1007/s11127-011-9865-0},
    publisher={Springer}
}

@article{BELANGER2008477,
	title = {Issue salience, issue ownership, and issue-based vote choice},
	journal = {Electoral Studies},
	volume = {27},
	number = {3},
	pages = {477-491},
	year = {2008},
	issn = {0261-3794},
	doi = {https://doi.org/10.1016/j.electstud.2008.01.001},
	author = {Éric Bélanger and Bonnie M. Meguid},
}

@article{FischerMichaela,
	author = {Fischer, Michaela},
	title = {From newspapers to social media? Changing dynamics in Swiss direct democratic campaigns},
	journal = {Swiss Political Science Review},
	volume = {29},
	number = {4},
	pages = {465-478},
	doi = {https://doi.org/10.1111/spsr.12578},
	year = {2023}
}

@article{ReveilhacMorselli,
	author = {Reveilhac, Maud and Morselli, Davide},
	title = {The Impact of Social Media Use for Elected Parliamentarians: Evidence from Politicians' Use of Twitter During the Last Two Swiss Legislatures},
	journal = {Swiss Political Science Review},
	volume = {29},
	number = {1},
	pages = {96-119},
	doi = {https://doi.org/10.1111/spsr.12543},
	year = {2023}
}

@article{ZumofenGuillaume,
	author = {Zumofen, Guillaume and Gerber, Marlène},
	title = {Effects of Issue-Specific Political Advertisements in the 2015 Parliamentary Elections of Switzerland},
	journal = {Swiss Political Science Review},
	volume = {24},
	number = {4},
	pages = {442-463},
	doi = {https://doi.org/10.1111/spsr.12333},
	year = {2018}
}

@article{GerberBühlmann,
	author = {Gerber, Marlène and Bühlmann, Marc},
	title = {Do Ads Add Up? The Impact of Parties' Advertisements on the Stability of Vote Choice at the Swiss National Elections 2011},
	journal = {Swiss Political Science Review},
	volume = {20},
	number = {4},
	pages = {632-650},
	doi = {https://doi.org/10.1111/spsr.12132},
	year = {2014}
}

\end{document}